\documentclass[reprint,aps,pre]{revtex4-1}

\usepackage{amsmath,amsfonts,amscd,amssymb,graphicx,subeqnar}
\usepackage[bbgreekl]{mathbbol}

\DeclareMathAlphabet{\mathpzc}{OT1}{pzc}{m}{it}
 
\usepackage{dcolumn}
\usepackage{overpic}
\usepackage{color}
\usepackage{adjustbox}
\usepackage{nomencl} 
\usepackage{etoolbox}
\makenomenclature

\DeclareSymbolFontAlphabet{\mathbb}{AMSb}
\DeclareSymbolFontAlphabet{\mathbbl}{bbold}

\renewcommand{\nomgroup}[1]{%
  \ifstrequal{#1}{L}{\item[\textbf{Latin Symbols}]}{%
  \ifstrequal{#1}{G}{\item[\textbf{Greek Letters}]}{%
  \ifstrequal{#1}{A}{\item[\textbf{Acronyms}]}{}}}}

\renewcommand{\vec}[1]{\mathbf{#1}}

\newcommand{\norma}[1]{\left|\left|{#1}\right|\right|}

\newcommand{\dpart}[2]{\frac{\partial {#1}}{\partial {#2}}}

\usepackage{hyperref}

\makeatletter
\@ifundefined{sf@counterlist}{\def\sf@counterlist{}}{}
\makeatother

\usepackage[dvipsnames]{xcolor}

\definecolor{bluePolimi}{RGB}{22, 44, 80}
\definecolor{lightBluePolimi}{RGB}{91, 122, 172}
\definecolor{redPolimi}{RGB}{180, 0, 0}
\definecolor{greenPolimi}{RGB}{78, 172, 91}
\definecolor{green2}{RGB}{0, 110, 0}
\definecolor{darkblue}{rgb}{0.0, 0.2, 0.6}
\definecolor{darkorange}{rgb}{0.8, 0.35, 0.0}

\makeatletter
\hypersetup{colorlinks=true}
\AtBeginDocument{\@ifpackageloaded{hyperref}
  {\def\@linkcolor{blue}
   \def\@anchorcolor{red}
   \def\@citecolor{red}
   \def\@filecolor{red}
   \def\@urlcolor{redPolimi}
   \def\@menucolor{red}
   \def\@pagecolor{cyan}
\begingroup
  \@makeother\`%
  \@makeother\=%
  \edef\x{%
    \edef\noexpand\x{%
      \endgroup
      \noexpand\toks@{%
        \catcode 96=\noexpand\the\catcode`\noexpand\`\relax
        \catcode 61=\noexpand\the\catcode`\noexpand\=\relax
      }%
    }%
    \noexpand\x
  }%
\x
\@makeother\`
\@makeother\=
}{}}
\makeatother

\usepackage{comment}
\usepackage{tabularx}
\usepackage{booktabs} 
\usepackage{parskip}

\begin{document}

\title{Constrained Sensing and Reliable State Estimation with Shallow Recurrent Decoders on a TRIGA Mark II Reactor}

\author{Stefano Riva$^{a, b}$, Carolina Introini$^{b}$, J. Nathan Kutz$^{a, c}$, Antonio Cammi$^{d,b,*}$}
\affiliation{$^{a}$Autodesk Research, 6 Agar Street, London, United Kingdom}
\affiliation{$^{b}$Politecnico di Milano, Department of Energy, CeSNEF - Nuclear Engineering Division, 20156 Milan, Italy}
\affiliation{$^{c}$Department of Applied Mathematics and  Electrical and Computer Engineering, University of Washington, Seattle, WA 98195}
\affiliation{$^{d}$Emirates Nuclear Technology Center (ENTC), Department of Mechanical and Nuclear Engineering, Khalifa University, Abu Dhabi, 127788, United Arab Emirates}
\email{antonio.cammi@polimi.it}

\begin{abstract} 
Shallow Recurrent Decoder networks are a novel data-driven methodology able to provide accurate state estimation in engineering systems, such as nuclear reactors. This deep learning architecture is a robust technique designed to map the temporal trajectories of a few sparse measures to the full state space, including unobservable fields, which is agnostic to sensor positions and able to handle noisy data through an ensemble strategy, leveraging the short training times and without the need for hyperparameter tuning. The architecture was successfully applied to the Molten Salt Fast Reactor concept; now, this work considers the performance of Shallow Recurrent Decoders on a deployed reactor concept. The underlying model is represented by a fluid dynamics model of the TRIGA Mark II research reactor; the architecture will use both synthetic temperature data coming from the numerical model and leveraging experimental temperature data recorded during a previous campaign. The objectives of this work are therefore: 1) presenting the first application of SHRED to a deployed nuclear reactor (TRIGA Mark II); 2) the integration of hybrid synthetic/experimental data within the SHRED framework; 3) a systematic assessment of SHRED robustness under physically constrained, low-dynamics sensor locations; and 4) a quantitative evaluation of SHRED self-correction capability when model-to-data discrepancies are present. Indeed, this approach is capable of accurately reconstruct every field of interest in real-time, using both synthetic (with an average relative error lower than 4\% in euclidean norm) and experimental data (with a RMSE of 1.52 K on the temperature, compared of 1.85 K of the CFD), making it suitable for interpretable monitoring and control purposes in the context of building digital twins for nuclear reactors.

\end{abstract}

\maketitle


\section{Introduction}

Understanding and reconstructing the internal state of nuclear reactors from sparse and noisy sensor measurements is a crucial task in reactor monitoring and control \cite{Leite2025-HTGRStateEstimation, karnik_constrained_2024}. Nuclear reactors are complex engineering systems featuring multi-scale and multi-physics phenomena, a harsh and hostile environment due to high temperatures and high fluence, especially for Generation-IV reactors \cite{cammi_data-driven_2024, aufiero2014development}: these facts pose several challenges for the state estimation tasks. Mathematically speaking, these tasks can be formulated as inverse problems where the system' state, represented by the main quantities of interest (e.g., temperature, power, velocity, pressure) over all the physical domain, is to be inferred starting from some sparse measurements of the state. This problem is typically ill-posed and it has to deal with 3 important challenges, valid for most engineering systems: (i) Sensors cannot be positioned arbitrarily due to physical access limitations, and their quantity is restricted by economic and instrumentation constraints \cite{argaud_sensor_2018}; (ii) there are some quantities which are not observable and their estimation can only be performed indirectly from other observable quantities \cite{INTROINI2023109538, gong_parameter_2023}, provided them to be coupled together; (iii) mathematical models, which are used to train and develop suitable techniques for state reconstruction, are affected by simplifying assumptions or uncertainty on the modelling parameters; therefore, state reconstruction methods should be able to fuse this background knowledge from the mathematical one with the one coming from real measurements on the system \cite{riva2024multiphysics}.

Data Assimilation (DA) is a general framework \cite{carrassi_data_2018, Cheng2023_MLDAReview} that aims at combining observational data with a mathematical background model to enhance the fidelity of predictions and facilitate the dynamic updating of model states \cite{Shutyaev2019aa, CHENG2025109359}. DA is strongly connected to the state estimation problem, which can be formulated in a variational sense: the optimal state is the one that minimises the distance between the background mathematical model and the sensor measurements \cite{carrassi_data_2018}. Traditional DA algorithms involve computationally heavy optimization problems or Kalman Filtering \cite{HU2025121046}, which can be infeasible for some applications where the prediction time should be very low (as in quasi-real-time). In particular, the main limitation is given by the computational costs associated to the solution of the background mathematical model, represented by a system of Partial Differential Equations, as for modelling nuclear reactor \cite{aufiero2014development, CERVI2019379, LUZZI20104873} or thermal-hydraulic facilities \cite{PINI2016108, RUIZ2015573, BATTISTINI2021116520, LUZZI2017262}. This is why some DA approaches are used in conjunction with Reduced Order Modelling (ROM) \cite{quarteroni2015reduced, lassila_model_2014, INTROINI2024113477, riva2024multiphysics}, such as the Proper Orthogonal Decomposition (POD) \cite{rozza_model_2020}, designed to extract the dominant features from the data and then work in a more compact and essential latent space. Nowadays, one of the most adopted approaches consists of fusing these compression methodologies with Machine Learning (ML) methods \cite{brunton_data-driven_2022, BruntonReviewML2020, Kobayashi2024aa}. The former allows obtaining a more essential data representation, thus significantly reducing the training times. ML approaches are becoming more and more interesting due to their intrinsic capabilities of learning a model directly from data and approximating non-linear functions \cite{goodfellow_deep_2016}. The combination of POD with Neural Network architectures is a growing field of research for fluid dynamics problems \cite{Nair_Goza_2020, Luo2023_PODAE, LI2025122003}, nuclear reactor problems \cite{gong_efficient_2022, gong_reactor_2024}, or energy and chemical engineering \cite{LEE2022118111, LIU2026122891, LIU2025121634}. Other promising strategies for ROM in the Deep Learning framework worth mentioning are the POD-DeepONet operator learning strategies \cite{LU2022114778}, the POD-DL-ROM architecture \cite{fresca_comprehensive_2021,fresca_pod-dl-rom_2022}, the Fourier Neural Operators \cite{dang2025flronetdeepoperatorlearning}, and the Voronoi tessellation \cite{Fukami:2021aa}. 

\begin{figure*}[tp]
    \centering
    \includegraphics[width=1\linewidth]{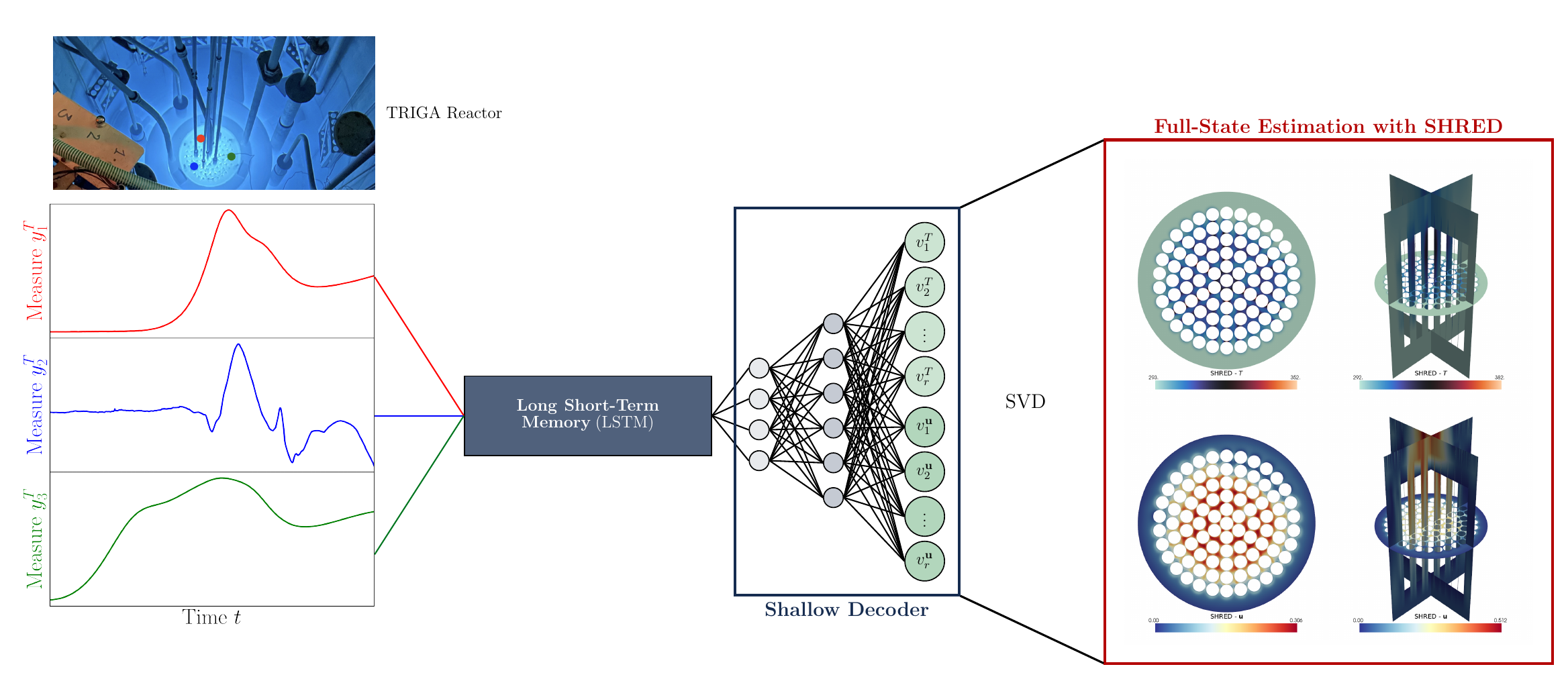}
    \caption{SHRED architecture for the TRIGA MARK II research reactor. Three sensors are used to measure the coolant temperature field. The measured time series is used to build a latent temporal sequence model to be mapped onto compressive representation, in time and space, of the field variables. These are then mapped to the full-order state space by singular value decomposition. This strategy permits laptop-level training in short times (order of minutes) even on multiple high-dimensional state vectors.}
    \label{fig: shred}
\end{figure*}

In this work, the focus will be on a novel and promising architecture capable of generating reliable reduced models by combining sparse measurements and the reduced dynamics, the SHallow REcurrent Decoder (SHRED) networks \cite{williams_sensing_2023, kutz_shallow_2024}. These architecture have been successfully applied to different physical problems, from biology to fluid dynamics and multi-physics simulations \cite{riva2024robuststateestimationpartial, shredrom, riva2025_parametricMSFR, riva_shreddynasty_2026, Rude2025}. Incorporating a recurrent unit and a shallow decoder, this neural network extracts a latent representation from sparse measurements to reconstruct the full state. SHRED leverages the Takens embedding theorem \cite{takens1981lnm} by substituting dense spatial data at a single instance with a time-trajectory of sensor measurements \cite{gao_sparse_2025}. Furthermore, the method lays its foundation in the theory of separation of variables for Partial Differential Equations (PDEs) \cite{williams_sensing_2023, shredrom}, separating temporal behaviour from the spatial structures. The application of SHRED to nuclear reactors and thermal-hydraulic systems has already been investigated in \cite{riva2024robuststateestimationpartial}, focusing on a single-parameter transient of the Molten Salt Fast Reactor and reconstructing the full state of 20 coupled fields, and in \cite{riva2025_parametricMSFR}, focusing on the parametric reconstruction of the full state of a Molten Salt Fast Reactor \cite{aufiero2014development}, starting from either out-core fast flux measures, in-core probes or sensors advected by the flow field; finally, in \cite{riva_shreddynasty_2026} the architecture has been applied to the DYNASTY experimental facility, adopting both synthetic and real measurements. In all cases, the results obtained were promising, thus motivating the continued investigation of this new architecture on different nuclear reactor systems. Despite the growing interest in data-driven state estimation for nuclear reactors, most existing approaches are validated exclusively on synthetic data from conceptual reactor designs. As mentioned above, the authors have investigated the use of real experimental data in \cite{riva_shreddynasty_2026}, though limited to 4 sensors only and considering a thermal-hydraulic experimental facility, but no prior work has applied a recurrent decoder architecture to a deployed research reactor using both synthetic and real experimental data, nor it has assessed its capability to self-correct model-to-data discrepancies under physically constrained sensing conditions. 

Specifically, this paper builds from all previous works and applies, for the first time, the architecture to a real reactor system, adopting both synthetic and experimental data: the work, in particular, focuses on the state estimation problem from temperature measurements in the TRIGA (Training Research Isotope production General Atomics) Mark II reactor \cite{international2016iaeaTRIGA} at the \textit{Laboratorio di Energia Nucleare Applicata} (LENA), University of Pavia. TRIGA is a 250 kW research reactor, mainly used for neutron activation analysis, education and training, and benchmarking to verify and validate numerical models \cite{ChiesaPhd}. A fluid dynamics model has been developed by Introini et al. \cite{INTROINI2023112118} using a Computational Fluid Dynamics (CFD) approach, which represents the background knowledge model. The objective of the paper is to investigate the reconstruction capabilities of SHRED during a heating transient from zero to full power, from limited sensors constrained in specific channels. Specifically, the sensors will be constrained to be selected in specific channels of the system with "low dynamics" to assess whether SHRED is truly robust and agnostic to sensor positioning so that even a non-optimal sensor location can provide the information needed to estimate the full system' state (composed of velocity, temperature, pressure, and turbulence quantities). Additionally, since experimental data at specific locations are available \cite{ChiesaPhd, INTROINI2023112118}, they will be used to assess the model self-correction/update capabilities of the architecture.

The novel contributions of this work are: (i) the first application of SHRED to a deployed nuclear reactor (TRIGA Mark II), as opposed to conceptual designs or small experimental facility; (ii) the integration of hybrid synthetic/experimental data within the SHRED framework; (iii) a systematic assessment of SHRED robustness under physically constrained, low-dynamics sensor locations; and (iv) a quantitative evaluation of SHRED self-correction capability when model-to-data discrepancies are present. The paper is organised as follows: Section \ref{sec: shred} presents the SHRED architecture; Section \ref{sec: triga-model} briefly describes the CFD model, along with some insights on the experimental setup; the key results are discussed in Section \ref{sec: num-res}; in the end, the main conclusions and future perspectives are drawn in Section \ref{sec: conclusions}.
\section{SHallow REcurrent Decoder} \label{sec: shred}

Shallow Recurrent Decoder networks were first proposed by Williams et al. \cite{williams_sensing_2023} as a promising sensing strategy with outstanding reconstruction capabilities in the limit of low data. SHRED has been further studied in \cite{ebers_leveraging_2023, kutz_shallow_2024, riva2024robuststateestimationpartial} and an in-depth discussion of its applicability to build Reduced Order Models has been presented in \cite{shredrom}. This architecture is particularly suitable for state estimation in nuclear reactors, where the number of sensors is limited due to physical, economic and safety constraints and the system is characterised by a high-dimensional state space \cite{riva2024robuststateestimationpartial, riva2025_parametricMSFR}. SHRED networks map measurement trajectories of an observable quantity to the full system state space, such as the complete thermal-hydraulic fields (temperature, pressure, velocity, and turbulence) of a nuclear reactor. The architecture comes with important advantages compared to other ML techniques. First, it is highly flexible regarding sensor configuration: in fact, sensors can be placed even randomly and usually limited to only $3\div5$. Second, training occurs in a compressed reduced space obtained with the Singular Value Decomposition (SVD) \cite{brunton_data-driven_2022}, thus it can be performed efficiently even on personal laptops, avoiding the need for powerful GPUs. Finally, and most importantly, SHRED exhibits minimal hyper-parameter tuning, as it has been shown how the same architecture can provide accurate results on a wide range of problems \cite{williams_sensing_2023, kutz_shallow_2024, shredrom, riva2024robuststateestimationpartial, riva2025_parametricMSFR, riva_shreddynasty_2026, Rude2025}. 

By handling inaccessible sensor locations and inferring unobservable fields, this methodology can support the creation of accurate nuclear digital twins \cite{mohanty_development_2021, zhang2025operatorlearningreconstructingflow}. The former is treated by agnosticism with respect to sensor positions; in fact, a state estimation can still be obtained accurately even with a few limited sensors, whereas the latter is supported by the intrinsic capability of neural networks to learn non-linear relationships. Most methods, proposed in the nuclear community \cite{argaud_sensor_2018, cammi_data-driven_2024, cannarile_novel_2018, karnik_constrained_2024, gong_empirical_2016}, covered the optimal sensor positioning for state estimation. Furthermore, even though good performance can be achieved with DA methods \cite{gong_phdthesis_2018, cammi_data-driven_2024}, the prediction capabilities of such methods are affected by the presence of constraints, i.e., regions in which sensors are not allowed to be placed \cite{maday_convergence_2016}.

The basic SHRED architecture is composed of a Long Short-Term Memory (LSTM) \cite{hochreiter1997long} and a Shallow Decoder Network (SDN) \cite{erichson2020shallow}: the former is used to capture the temporal dynamics of the sparse measurements, while the latter is used to decode the latent space to the full state space. The hyperparameters for each component are the same of the original work of Williams et al. \cite{williams_sensing_2023}, they are unchanged and according to authors' experience they represent an optimal configuration applied to different problems \cite{riva2024robuststateestimationpartial, shredrom, riva2025_parametricMSFR, riva_shreddynasty_2026}: both components are composed of 2 hidden layers, with the former having 64 neurons per layer and the latter consisting of 350 and 400 neurons, respectively.

As previously said, SHRED is based on variable separation, generalised to neural network approximations. In addition to this fundamental mathematical concept, the Takens embedding theorem \cite{takens1981lnm} is used to justify how a latent representation can be generated by time-delayed embeddings of the original data within the LSTM. A theoretical analysis of these mathematical foundation concepts can be found in \cite{shredrom}. Let $\vec{u}_k=\vec{u}(t_k)\in\mathbb{R}^{\mathcal{N}_h}$ be the high-dimensional state at time $t_k$ and let $\vec{y}^s_k=\mathbb{H}\vec{u}(t_k)\in\mathbb{R}^{s}$ be some sparse sensor measurements, where $\mathbb{H}\in\mathbb{R}^{s\times\mathcal{N}_h}$ is the observation operator \cite{karnik_constrained_2024}. SHRED has been conceived to reconstruct the high-dimensional state space starting from few sparse sensors, but, instead of mapping the state space $\vec{u}(t_k)$ from a single set of $s$ measurements $\vec{y}^s(t_k)\in\mathbb{R}^s$ at time $t_k$, as in \cite{Nair_Goza_2020, Luo2023_PODAE, zhang2025operatorlearningreconstructingflow}, the temporal history of the measurements is exploited to reconstruct the full state space. During the offline phase, a collection of snapshots of the high-dimensional state and some sparse measures are required for $k=1, \dots, N_t$. The measurements are divided into sequences of different trajectories, of length $L$, through the time-delay embedding process: each one will be $[\vec{y}^s_{k}, \dots, \vec{y}^s_{k-L}]$ for $k=1, \dots, N_t$ and with pre-padding applied \cite{shredrom} to obtain the state estimation starting from the beginning of the transient and avoiding the blind period before the $L$-th time step.

The size $\mathcal{N}_h$ of the high-dimensional state $\vec{u}_k$ can be very large, e.g., $\mathcal{N}_h\sim 10^{4\div 6}$, thus making the training of the neural network computationally very expensive. To overcome this issue, it becomes convenient to reason in a compressed space \cite{kutz_shallow_2024}: instead of dealing with the high-dimensional data, the snapshots are projected into a low-dimensional space using the SVD/POD. Let $\mathbb{X}\in\mathbb{R}^{\mathcal{N}_h\times N_t}$ be the matrix containing the high-dimensional states, in each column, and let $\mathbb{U}\in\mathbb{R}^{\mathcal{N}_h\times r}$ be the matrix containing the SVD/POD modes of $\mathbb{X}$, with $r$ as the number of modes to be retained \cite{rozza_model_2020}. A compressed representation is found by projection, i.e.:
\begin{equation}
    \vec{v}_k = \mathbb{U}^T\vec{u}_k \quad \longleftrightarrow \quad 
    \mathbb{V} = \mathbb{U}^T\mathbb{X}\in\mathbb{R}^{r\times N_t}
\end{equation}
The training process, now called \textit{compressive training}, is then performed in the compressed space \cite{shredrom}, ensuring a more efficient procedure. In this way, the SHRED does not have to map the measurements to the high-dimensional state; rather, it focuses on the relation between $\vec{y}^s_k$ and $\vec{v}_k$. Let $\vec{f}_R$ be the function representing the recurrent neural network, i.e., the LSTM, encoding the measures over a time window $[t_{k-L}, t_k]$ into a latent representation $\vec{z}_k$ of dimension $64$, i.e.:
\begin{equation}
    \vec{z}_k = \vec{f}_R(\vec{y}^s(t_{k}),\vec{y}^s(t_{k-1}), \dots, \vec{y}^s(t_{k-L}))
\end{equation}
in which the lagging value $L$ is a hyperparameter that must be tuned properly according to the problem under study, according to the authors' experience, accurate results can be obtained with $L\sim 30$. Then, the reduced state space ${\vec{v}}(t_k)=\vec{v}_k$ is reconstructed by the SDN, being $\mathbf{f}_D$ the function representing the decoder network, by mapping the latent representation $\vec{z}_k$ to, respectively, the reduced space, i.e.:
\begin{equation}
    \hat{\vec{v}}(t_k) = \vec{f}_D(\vec{z}_k) = \vec{f}_D(\vec{f}_R(\vec{y}^s(t_{k}), \dots, \vec{y}^s(t_{k-L}))
\end{equation}
finally, the high-dimensional state is retrieved by multiplying this output by the SVD modes, i.e., $\hat{\vec{u}}(t_k)=\mathbb{U}\hat{\vec{v}}(t_k)$.

After splitting the available input-output data pairs into train $\Xi_{\text{train}}^t$, validation $\Xi_{\text{valid}}^t$ and test $\Xi_{\text{test}}^t$, according to the purpose of the state estimation (which can be either reconstruction, prediction or forecast or parametric reconstruction), the recurrent network $\vec{f}_R$ and the decoder network $\vec{f}_D$ are trained by minimizing the reconstruction error on the reduced state:
\begin{equation}
\begin{split}
    {\mathcal{J}} &= \sum_{k\in\Xi^t_{\text{train}}} \norma{\mathbb{U}^T\vec{u}_k - \hat{\vec{v}}_k}_2^2 \\
    &= \sum_{k\in\Xi^t_{\text{train}}} \norma{\mathbb{U}^T\vec{u}_k - \vec{f}_D\left(\vec{f}_R\left(\vec{y}^s_{k}, \dots , \vec{y}^s_{k-L}\right)\right)}_2^2
\end{split}
\label{eqn: shred-loss-compr}
\end{equation}

SHRED is agnostic to sensor positions, meaning that if the measurements contains sufficient statistical significance then the position of the sensor is irrelevant \cite{kutz_shallow_2024, riva2024robuststateestimationpartial}. This aspect can be further exploited united with the fact that SHRED can work with few sensors as input: the authors have shown in \cite{riva2024robuststateestimationpartial, riva2025_parametricMSFR} that SHRED can operate in ensemble mode to better handle noisy data. If more than 3 sensors are available in the facility, instead of training a single SHRED model using all the available measurements, it is more convenient to train a several SHRED models. In this way, the output of each model can be ensembled obtaining a mean of the prediction and an associated confidence interval (further discussion on this topic are available in \ref{sec: ensemble-std}). Finally, SHRED can be relevant to the regulation and licensing of advanced reactors. The architecture enables accurate monitoring during nominal\footnote{Nominal refers to standard design operating conditions, specifically rated power operation (250 kW) with natural circulation cooling active and no anomalous transient in progress.}  operating conditions, as demonstrated in this work; its applicability to accidental scenarios is supported by previous studies on the Molten Salt Fast Reactor \cite{riva2025_parametricMSFR,riva2024robuststateestimationpartial} and is a subject of ongoing investigation for the TRIGA system. As said above, SHRED provides confidence intervals, ensuring robust predictions especially with noisy measures. Furthermore, its rapid training capability facilitates offline deployment, mitigating the risk of cyber attacks associated with online systems.

\section{TRIGA Mark II Research Reactor}\label{sec: triga-model}

\begin{figure*}[tp]
    \centering
    \begin{overpic}[width=0.48\linewidth]{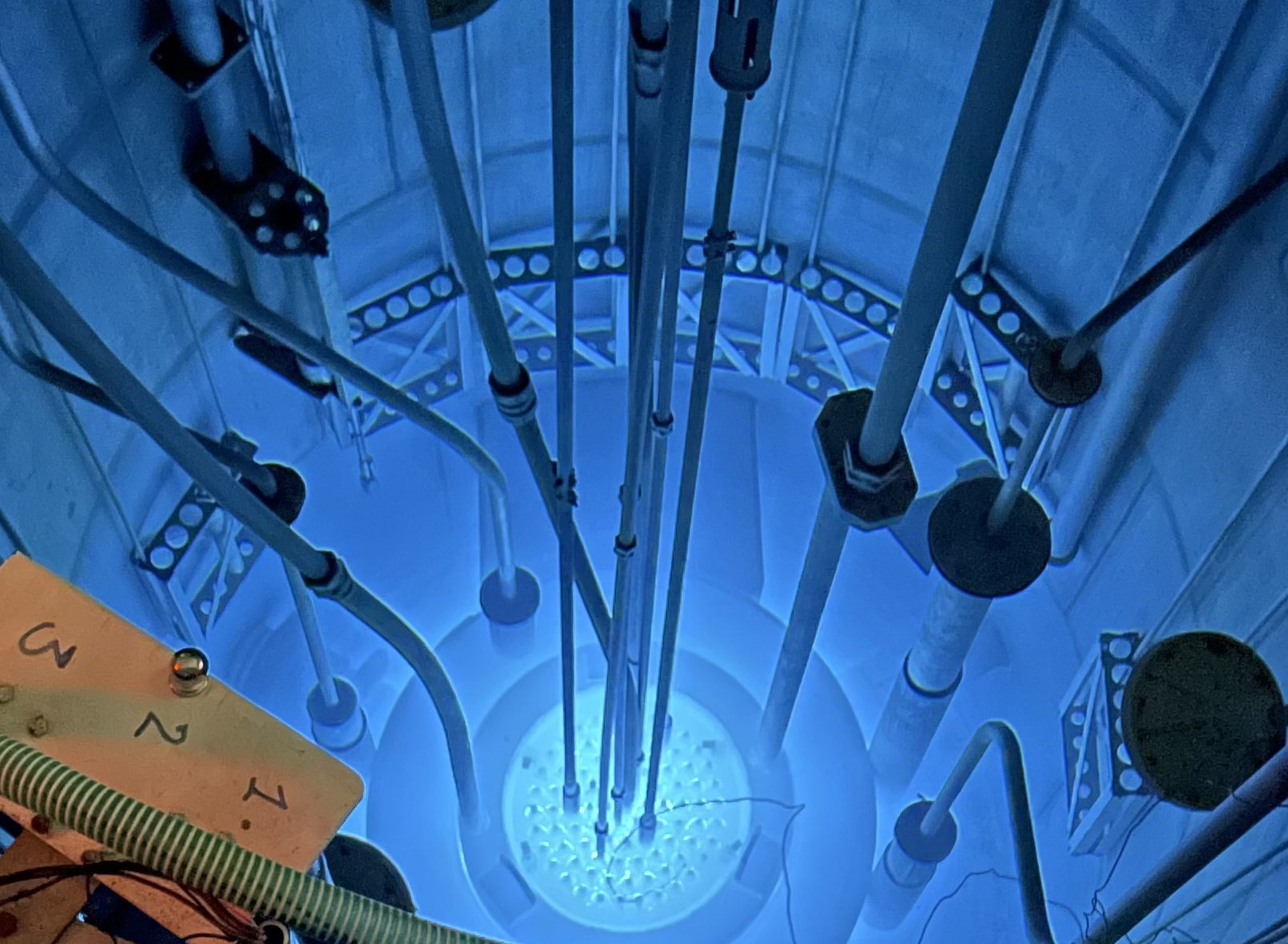}
    \put(47,-5){(a)}
    \end{overpic}
    \begin{overpic}[width=0.43\linewidth]{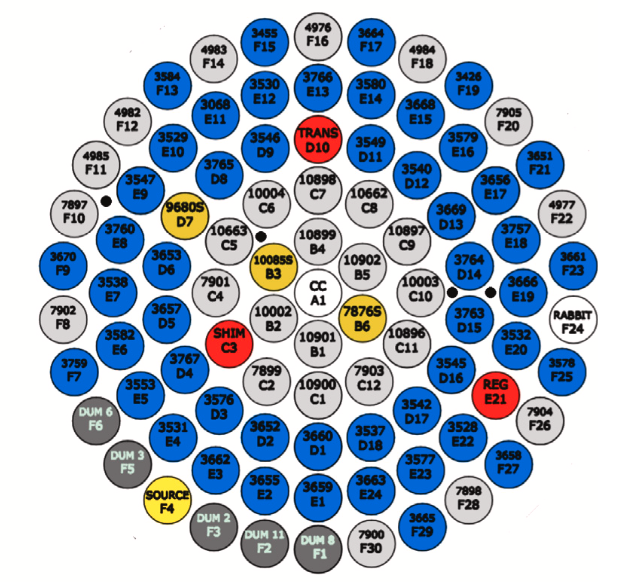}
    \put(48,-5){(b)} 
    \end{overpic}
    \caption{(a) Photo of the Cherenkov effect at the TRIGA reactor in Pavia, visible after the reactor shutdown following SCRAM; (b) Disposition of the fuel elements (blue with aluminium and light grey with steel cladding) in the current core configuration. Red elements are the control rod, white ones the irradiation channels, dark grey ones the graphite elements, the yellow one the neutron source, and the black dots are the locations of the available coolant temperature measurements.}
    \label{fig: triga}
\end{figure*}

Innovative methods in the world of scientific machine learning require proper benchmarks to analyse their performance, and research reactors offer interesting opportunities, since the most significant phenomena observed in commercial plants are also present in research reactors. Compared to large-scale commercial systems, high-fidelity simulations of these reactors are typically less expensive, allowing for verifying and testing numerical models; furthermore, experimental data are available, allowing for validation of the proposed methodologies and models before their use for commercial reactors. For these reasons, this work adopts the TRIGA Mark II reactor (Training, Research, Isotopes, General Atomics) at the University of Pavia, a 250 kW pool-type reactor cooled through natural circulation \cite{international2016iaeaTRIGA}, shown in Figure \ref{fig: triga} along with the current core configuration. This pool reactor is cooled by light water through natural convection; for long-time operation, an auxiliary external forced cooling can be activated if the pool temperature reaches the threshold value of 40 $^o$C. The reactor presents 91 fuel locations, disposed in six concentric rings (Figure \ref{fig: triga}b): the configuration adopted in this work refers to the one adopted in September 2013 \cite{INTROINI2023112118}, with 80 fuel elements mounted. The core features a lower and upper grid acting as support of the core weight (the former) and of the fuel elements, guaranteeing correct spacing (the latter). In the end, the reactor is surrounded by a graphite reflector. \ref{app: model} briefly presents the OpenFOAM model used to generate the data: interested readers can refer to  \cite{introini_advanced_2021, INTROINI2023112118} for a complete description of the core and the main geometrical parameters.

Furthermore, the INFN (\textit{Istituto Nazionale di Fisica Nucleare}, National Institute of Nuclear Physics) performed an experimental campaign in the first half of 2016, providing temperature measurements which have been used to validate the CFD model in \cite{INTROINI2023112118} and adopting the experimental setup developed and installed in the reactor in 2013 \cite{ChiesaPhd}. In particular, sensors (Pt1000 Resistance Temperature Detectors) for the coolant temperature are placed in specific channels of the reactor core\footnote{See Figure 5 from \cite{INTROINI2023112118}, note that the core configuration for the fuel elements reported there refers to the first configuration, not to the one used here}, they are mounted on aluminium rods, each with eight detectors equally spaced in the axial direction (at a distance of 8 cm one from the other) and labelled TC1 to TC8 from bottom to top (the bottom one is located 8 cm from the top of the lower grid). The experimental data collected refers to nominal operating conditions (reactor at 250 kW) following a power transient from zero power, with forced cooling off. Each sensor has a reported uncertainty of 0.1 $^o$C. Other measurement uncertainties may be present, referring to the electronic noise of the data acquisition system and the exact positioning of the aluminium rods: specifically, it is impossible to verify whether the rod remains in position during the reactor operation. Interested readers should refer to \cite{ChiesaPhd, INTROINI2023112118} for more details on the experimental campaign.

\section{Numerical Results}\label{sec: num-res}

This work discusses the application of SHRED to the CFD model of \cite{introini_advanced_2021} (and briefly reported in Section \ref{sec: triga-model}) of the TRIGA Mark II reactor. Two different cases will be investigated:
\begin{enumerate}
    \item Synthetic case only, using SHRED: two channels characterised by low-dynamics (one outside the external fuel ring and one of the cooling channels behind a control rod, thus "hidden") are the only available regions where sensors can be placed.
    \item Using experimental data for testing SHRED, assessing not only its capability of creating accurate reduced-order models but also its capability of self-updating itself when real data (unavoidably presenting some difference compared to the model estimate) are employed.
\end{enumerate}

For all the following cases, the state vector $\mathcal{V}$ is composed by temperature $T$, velocity $\vec{u}$, normalised pressure $p$, turbulence kinetic energy $\kappa$ and dissipation rate $\omega$, i.e. $\mathcal{V} = [T, \vec{u}, p, \kappa, \omega]$. The only observable field is the temperature in specific locations. From the CFD model in Appendix \ref{app: model} and \cite{introini_advanced_2021}, $N_t = 500$ snapshots are available: the data have been randomly split into train, validation and test sets with ratio 70\%-15\%-15\% in reconstruction mode (meaning that no prediction or forecast outside the training range is foreseen). Before training, the data have been normalised using a \textit{min-max} scaler from scikit-learn \cite{scikit-learn}: this choice guarantees a more stable compression and avoids working with the large differences in the magnitudes since different fields have quite different scales. Finally, the SVD for each field in $\mathcal{V}$ is performed to obtain the modes $\mathbb{U}$ and the coefficients $\mathbb{V}$. 

\begin{figure}[tp]
    \centering
    \includegraphics[width=0.75\linewidth]{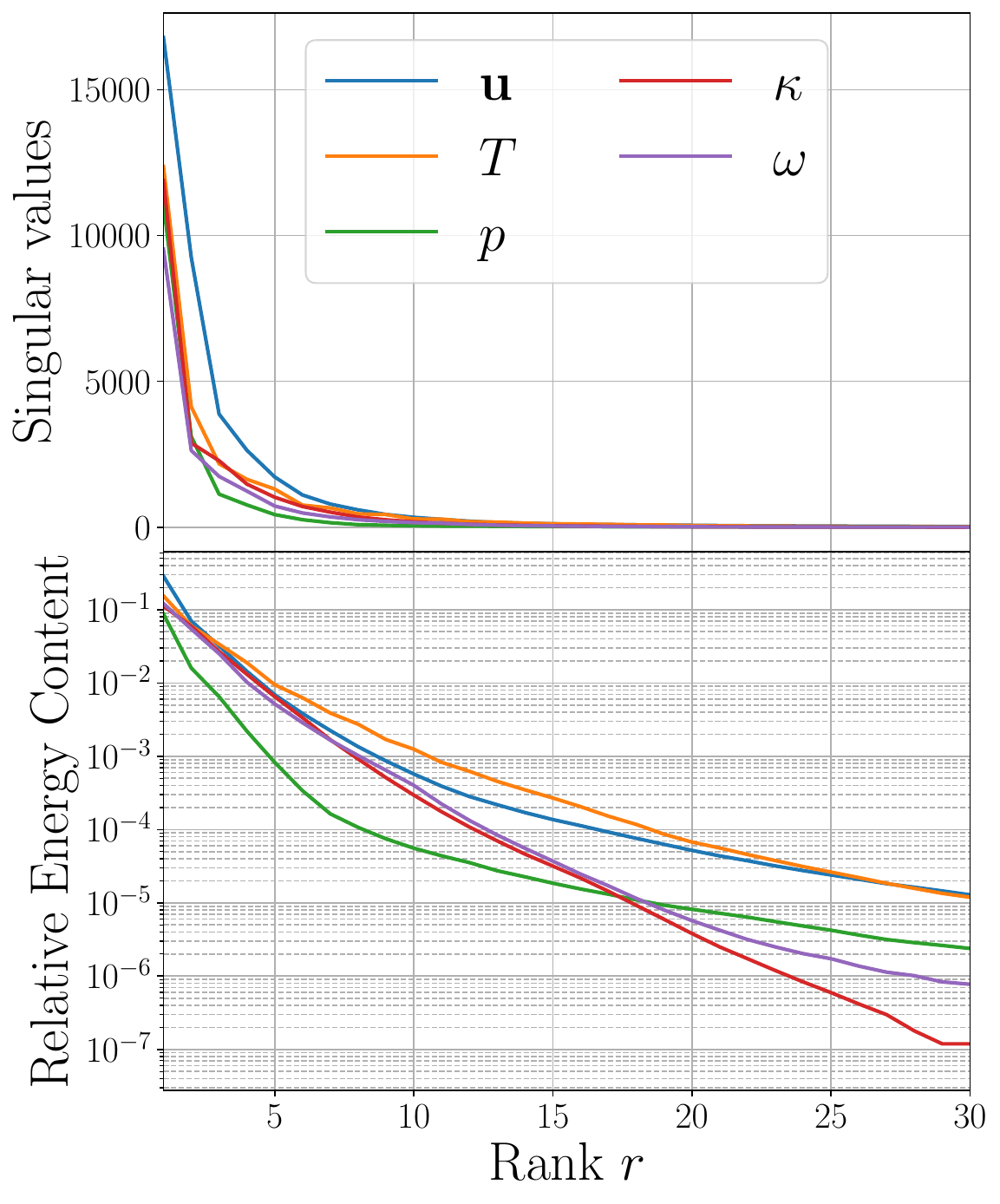}
    \caption{Singular values decay (top) and relative energy content discarded by the SVD modes (bottom) for the fields in the state vector $\mathcal{V} = [T, \vec{u}, p, \kappa, \omega]$ for the TRIGA Mark II reactor.}
    \label{fig: triga-svd}
\end{figure}

A fundamental preliminary analysis is the assessment of the reducibility of the problem by looking at the singular values decay of the SVD (Figure \ref{fig: triga-svd}) and the residual energy/information content \cite{rozza_model_2020}. From the plot, it emerges how a rank $r=15$ is sufficient to retain 99.9\% of the total energy; for velocity, this implies that most of the large scales are collected and encoded in the SVD modes. Note that for any field in $\mathcal{V}$, a compression via SVD has to be performed to obtain for each field a specific representation through the basis: a single shallow decoder has been used to map the latent space of the recurrent unit to the full reduced state to limit the number of parameters to learn during the neural network training. For some applications, it may be useful to train a sequence of decoders whose length is equal to the number of fields as in \cite{kutz_shallow_2024}.

\subsection{Reconstruction of the state from constrained sensors with poor dynamics}\label{sec: core-rec-synthetic}

\begin{figure}[tp]
    \centering
    \includegraphics[width=0.75\linewidth]{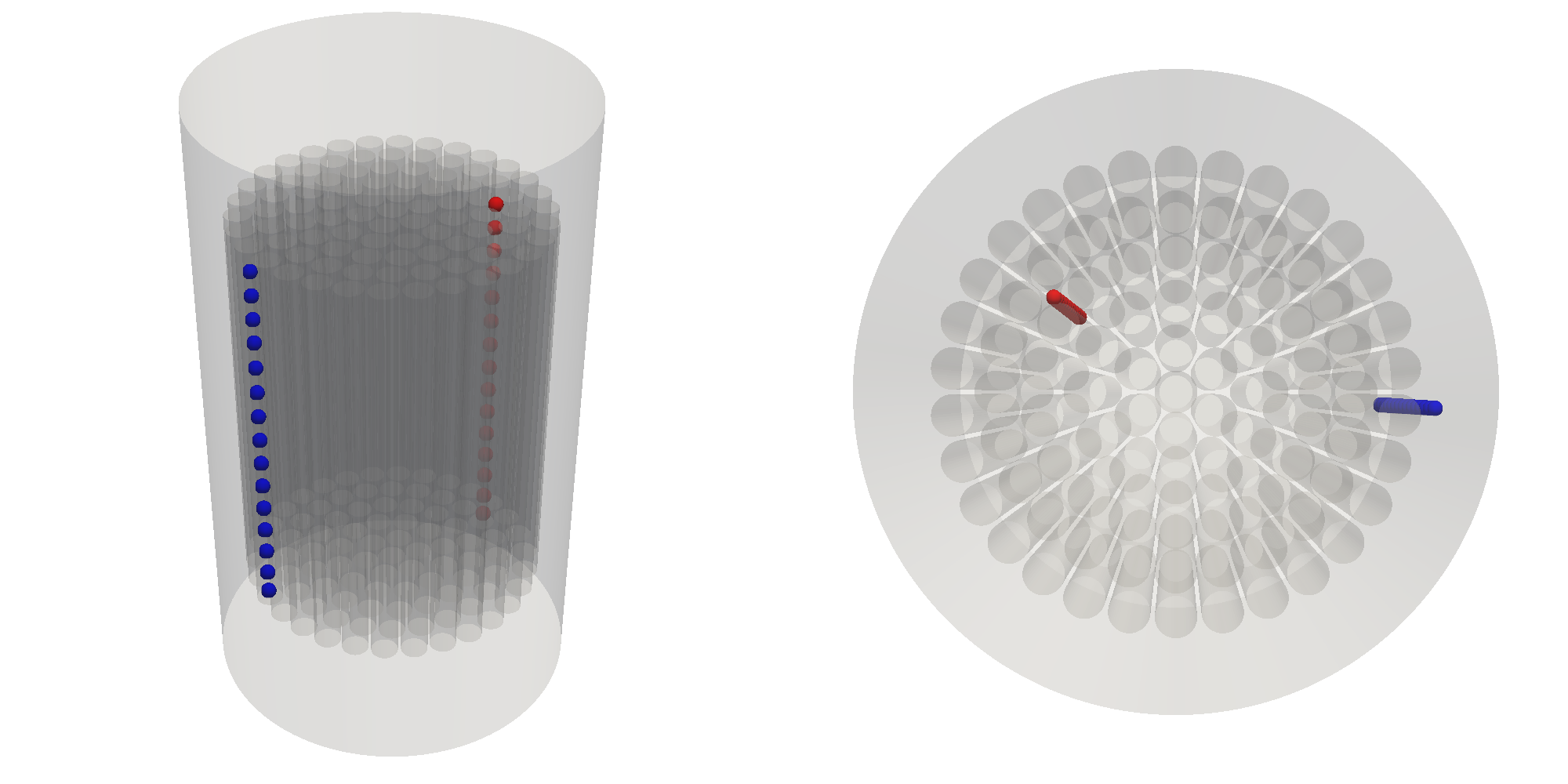}
    \caption{Computational geometry of the TRIGA reactor (shaded in gray): the available positions for sensing are the red and blue channels, being the REG and EXT respectively.}
    \label{fig: sensors-channels-core}
\end{figure}

This section investigates the use of SHRED to reconstruct the full state space of the CFD model of the TRIGA Mark II reactor core, starting from sparse temperature sensors only (as in the real reactor, this is the only field which can be measured and hence this choice is made to mimic reality), \textit{a-priori} constrained to be in two specific channels of the reactor core, as shown in Figure \ref{fig: sensors-channels-core}: these two channels are \textit{ext} (blue), representing a peripheral channel outside the most external fuel ring where the dynamics are less pronounced, and \textit{reg} (red), the cooling channel shielded by the REG control rod. 20 available positions within the reactor core are chosen in the computational model, these are equispaced along the axial direction.

The sensors are modelled as linear functionals with a Dirac delta as kernel, so that the measures $\vec{y}^s\in\mathbb{R}^3$ extracted from the CFD model,  for $k=1,2,3$, are defined as:
\begin{equation}
  y_k^{s,T}(\cdot) = (1+\epsilon)\cdot  \int_\Omega T(\vec{x}; \cdot) \cdot \delta (\vec{x} - \vec{x}_k)
    \label{eqn: chap07-triga-sensor-def}
\end{equation}
given $\epsilon$ as the random noise, assumed to be Gaussian with standard deviation $\sigma$, $\Omega$ as the spatial domain and $\vec{x}\in\Omega$ as the space vector. In particular, the state estimation will be evaluated in accuracy and robustness using synthetic data with constrained sensors and noisy measurements.

\begin{figure*}[tp]
    \centering
    \includegraphics[width=1\linewidth]{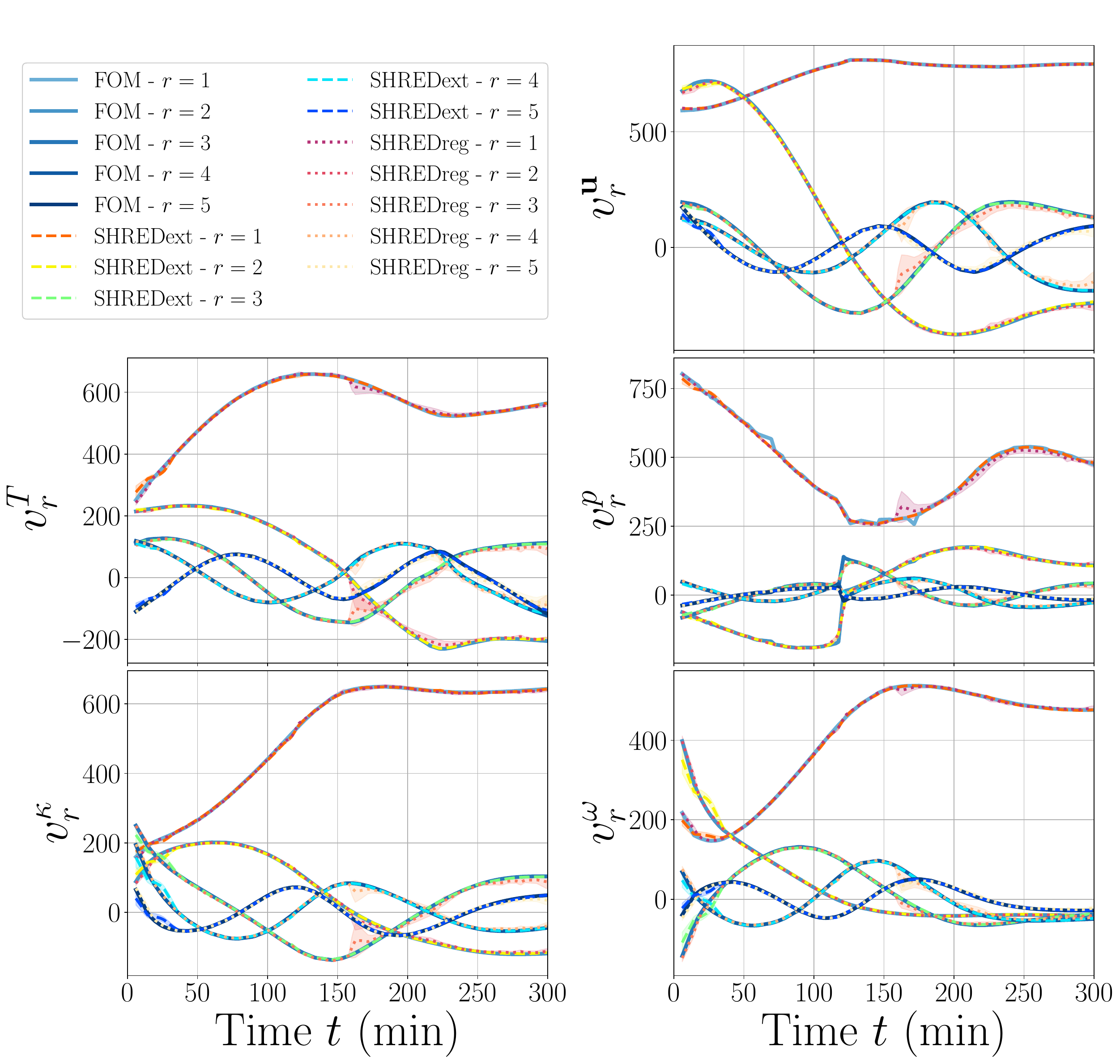}
    \caption{SHRED reconstruction of the first 5 SVD coefficients of velocity $\vec{u}$, temperature $T$ (observed field), pressure $p$, turbulent kinetic energy $\kappa$ and dissipation rate $\omega$, for test time only. The mean of the SHRED models is reported with dashed curves (\textit{ext} channels) and dotted curves (\textit{reg} channels), whereas the continuous lines are the reference from the full-order data; the shaded areas indicates the uncertainty of the SHRED ensemble.}
    \label{fig: chap07-triga-red-dyn-rec}
\end{figure*}

The ensemble strategy for SHRED (see Section \ref{sec: shred}) has been adopted to provide a more robust estimation and reconstruction of the state of the reactor over time. In particular, for each channel, $\mathcal{L}=10$ different sensor configurations have been used (following the discussion in \cite{riva2025_parametricMSFR}), sampling from the 20 available channel positions a subset of 3 sensors used to measure the temperature field and later train the architecture (Figure \ref{fig: sensors-channels-core}). The measures synthetically generated from these sensors are polluted by random noise as in Eq. \eqref{eqn: chap07-triga-sensor-def}: in particular, the temperature $T$, scaled to [0,1], is polluted with random Gaussian noise $\epsilon$ with standard deviation $\sigma = 0.025$ corresponding to about $\pm5$ K of uncertainty on the original data. 

Then, all the SHRED models are trained: a single SHRED requires less than a minute for training on a MacBook Pro M4 Pro (\texttt{mps} device). The reconstructed reduced state space from each case (\textit{ext} and \textit{reg}) is compared with the ground truth in Figure \ref{fig: chap07-triga-red-dyn-rec} for all the fields in $\mathcal{V}$. The continuous lines represent the \textit{ground-truth}, i.e., the reduced coefficients obtained by directly projecting the snapshots; the dotted and dashed curves are the mean prediction of SHRED\footnote{Only the prediction at test indices is reported. Due to the smoothness of the POD coefficients the test elements are sufficient to properly visualise the function in a continuous way.} from \textit{ext} and \textit{reg} channels, respectively; the shaded area is the standard deviation associated with the SHRED estimation. For all quantities, including the non-measured ones, there is a very good agreement between the prediction and the true value, highlighting how SHRED can learn the reduced dynamics from sparse sensors placed in "bad" (in the sense of observed dynamics) channels of the TRIGA reactor.

Once the reduced dynamics have been learnt, it must be assessed how the actual fields are predicted at the high-dimensional level by reconstructing the SHRED output using the decompression from SVD. In particular, the average relative error $\varepsilon_2^\psi$ for the generic variable $\psi$ is defined as:
\begin{equation}
    \varepsilon_2^{\psi} =  \frac{1}{N_t}\sum_{j=1}^{N_t}\frac{\norma{\boldsymbol{\psi}_{j} - \hat{\boldsymbol{\psi}}_{j}}_2}{\norma{\boldsymbol{\psi}_{j}}_2}
    \label{eqn: error-nonparametric-shred}
\end{equation}

\begin{figure}[tp]
    \centering
    \includegraphics[width=0.75\linewidth]{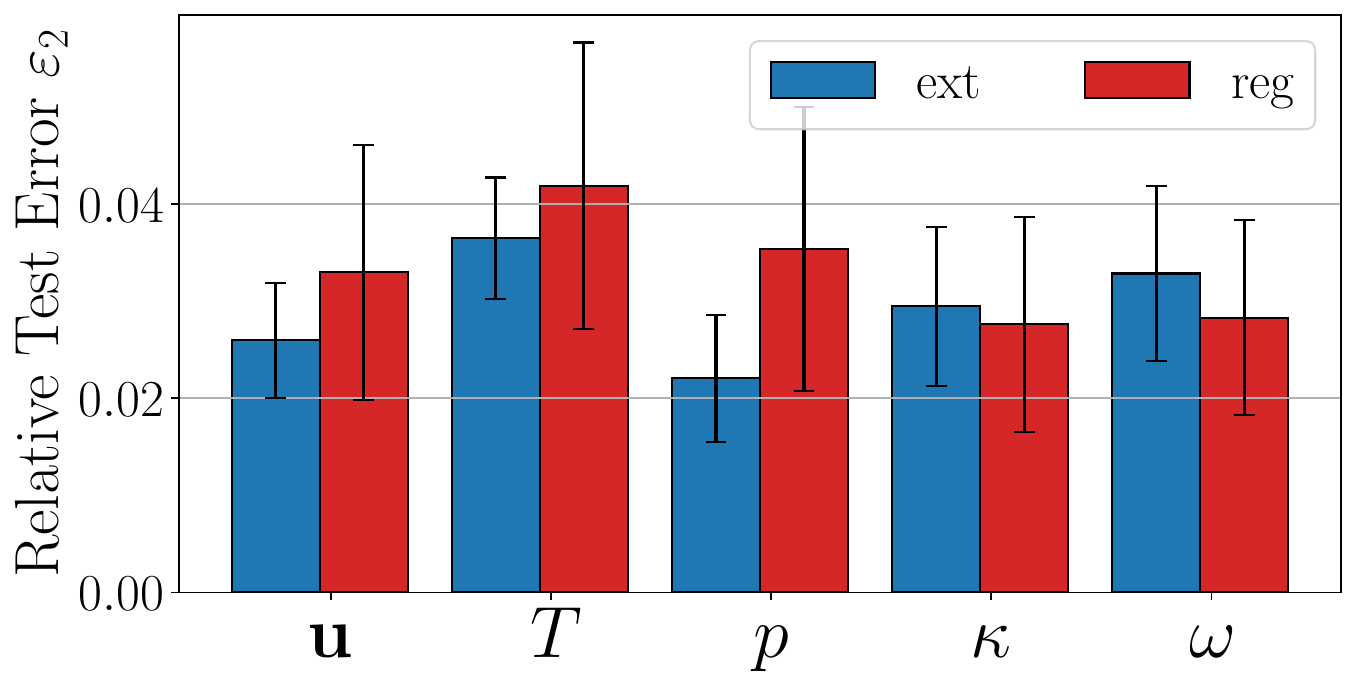}
    \caption{Average (of all configurations) relative errors $\varepsilon_2$ between SHRED reconstruction and FOM, for the test set measured in the $l_2$-norm, using sensors either in the \textit{ext} or \textit{reg} channel.}
    \label{fig: chap07-triga-fomerrors}
\end{figure}

The average relative test error, Eq. (6), for each field is plotted in Figure \ref{fig: chap07-triga-fomerrors} for the cases in which sensors are placed either in the \textit{ext} (blue) and \textit{reg} (red) channels: this quantity serves as a global correctness indicator for the method. The performance are pretty similar, except for the turbulent quantities. The reason for this behaviour can be explained by the fact in the external channel the flow is almost laminar and hence less information about the turbulent quantities is given to the system; nevertheless, overall the mean errors are all under 2\%, meaning that a very good agreement between the SHRED prediction and the FOM is reached.

\begin{figure*}[tp]
    \centering
    \includegraphics[width=1\linewidth]{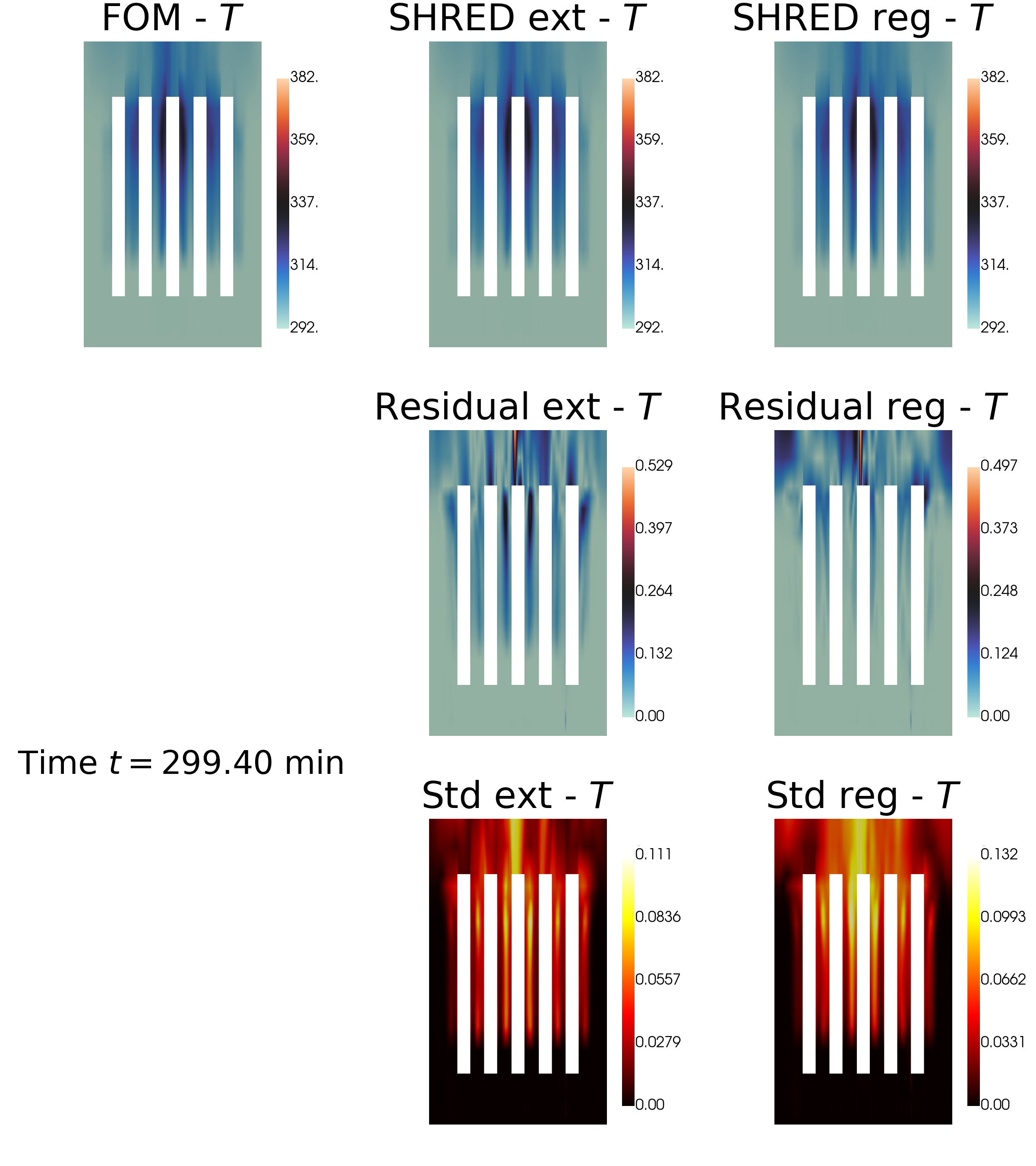}
    \caption{Contour plots for the last time instant in the test set of the observable field $T$: the section at $y=0$ of the reactor core is displayed. From left to right on the first row: full-order solution, mean of the SHRED ensemble for \textit{ext} and \textit{reg} and the associated standard deviation. The second row is the residual field and the last one the standard deviation associated to the prediction.}
    \label{fig: chap07-triga-contours-recon-T}
\end{figure*}

\begin{figure*}[tp]
    \centering
    \includegraphics[width=1\linewidth]{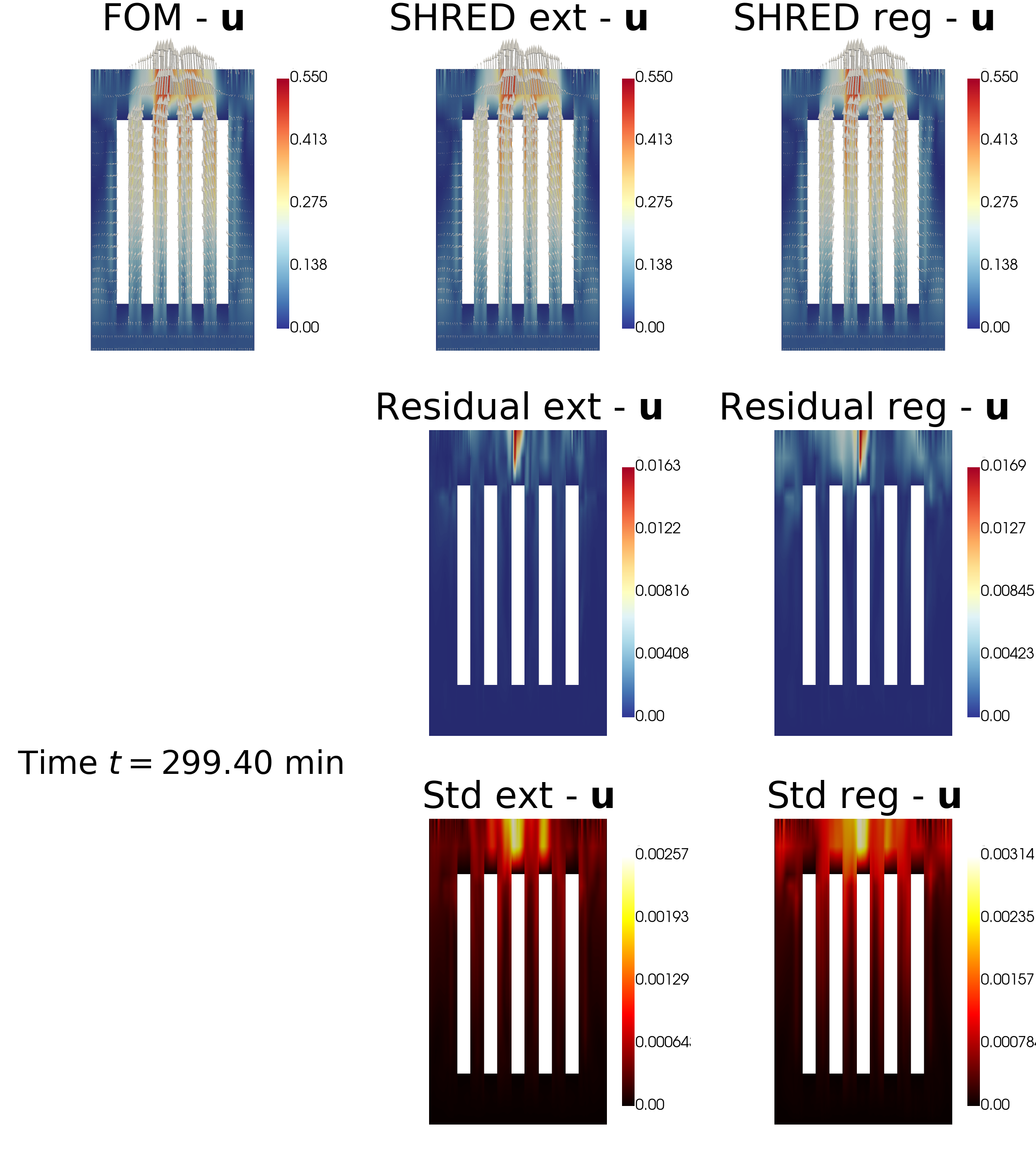}
    \caption{Contour plots for the last time instant in the test set of the unobservable field $\vec{u}$: the section at $y=0$ of the reactor core is displayed. From left to right on the first row: full-order solution, the mean of the  SHRED ensemble for \textit{ext} and \textit{reg} and the associated standard deviation. The second row is the residual field and the last one the standard deviation associated to the prediction.}
    \label{fig: chap07-triga-contours-recon-U}
\end{figure*}

Figure \ref{fig: chap07-triga-contours-recon-T} and \ref{fig: chap07-triga-contours-recon-U} shows contour plots, taken at the xy mid-plane, of the SHRED prediction at the last time step (of the test set), compared against the FOM (temperature and velocity respectively) for the case with sensors in \textit{ext} (2$^{\text{nd}}$ column) and \textit{reg} (3$^{\text{rd}}$ column); the associated standard deviation of the reconstructed fields is also shown, allowing to see the zones with higher variation and hence pinpointing the dominant structures cut off by the compression. For the temperature field (Figure \ref{fig: chap07-triga-contours-recon-T}), for the case with sensors placed in the \textit{reg} channel there is an uncertainty of about $\pm0.3$, whereas for the \textit{ext} channel the uncertainty value is lower. This difference could be a hint that the measurements from the latter channel are more statistically significant and, therefore, allow achieving a more reliable prediction: from the physical stand point, the \textit{reg} channel is deeply affected by the presence of the control rod, which absorbs neutrons and hence reduces the local power, which produces a temperature field not fully representative of the overall system. However, the performance of SHRED is quite good in both cases, with a very accurate reconstruction of the entire full state; despite slight differences, the sensors positions does not affect too much the performance of SHRED.

\subsection{Model Update with Experimental Data}

The SHRED architecture has been proven to be a powerful tool to reconstruct the reactor' state using synthetic data for cross-validation; however, the main goal of these algorithms consists in adopting them for the monitoring and control of real reactor dynamics; hence, the performance when real experimental data are considered (for now, focusing still on reconstruction mode) must be evaluated. Local observations of quantities provide a true insight into the phenomena occurring inside the system, and they represent the best knowledge of the reality itself. Inevitably, any model adopted to describe a physical system will show some discrepancy with experimental data due to assumptions made or uncertainty on physical parameters: as mentioned in \cite{riva2024multiphysics}, data-driven ROM approaches should be able to update the background knowledge of the model used for training to include the additional information provided by the measurements themselves, as can be done using 'classical' data-driven ROM methods such as the Generalised Empirical Interpolation Method \cite{maday_generalized_2015} and Parameterised-Background Data-Weak formulation \cite{maday_parameterized-background_2014}. In this framework, this section will investigate what is the amount of update/correction the architecture can provide when real experimental data are available during the online phase.

For the present analysis, among the instrumented channels, two of them have been selected (Hole 3 and Hole 9 as per Figure 5 from \cite{INTROINI2023112118}), whose experimental measurements are used for testing the SHRED model and its capabilities for correcting/updating the reconstruction. Two different setup will be considered: from the eight available locations of Hole 3, $\mathcal{L}=20$ subsets have been extracted (from a maximum number of possibilities of 56) to train the SHRED models with CFD data only; then, the experimental data from both channels are used to test the architecture once it has been trained and deployed; similarly, the second case considers as available positions for training those in Hole 9 and the experimental measures from both channels are used for testing. This first analysis aims to investigate if there are different correction effects when the test input data are not from the "same dataset" of the training, namely from the same synthetic simulation, and assess if SHRED can work with real experimental data for monitoring and control.

\begin{figure*}[tp]
    \centering
    \begin{overpic}[width=0.95\linewidth]{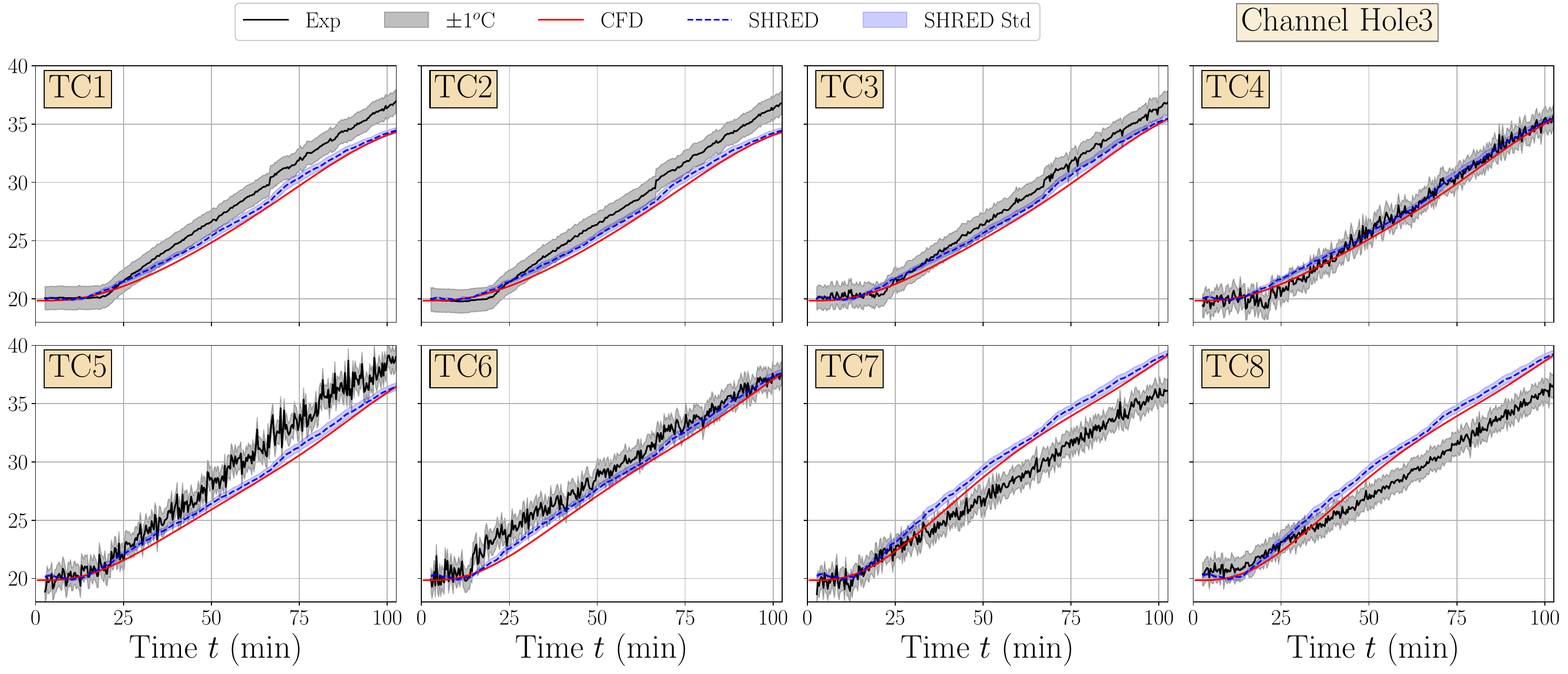}
    \put(95,41){(a)}   
    \end{overpic}\vspace{0.5cm}
    \begin{overpic}[width=0.95\linewidth]{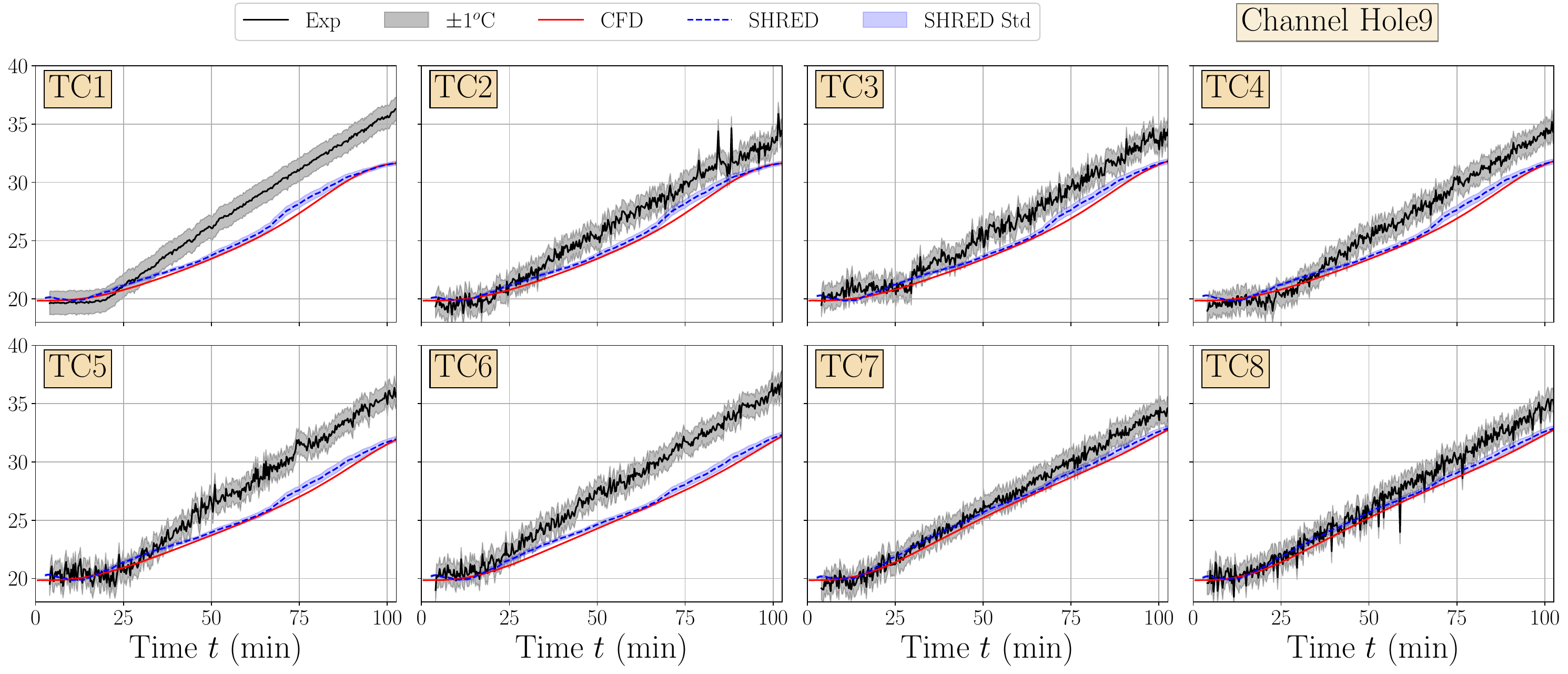}
    \put(95,41){(b)}   
    \end{overpic}
    \caption{Dynamical evolution of the temperature at the experimental locations in Hole 3 (a, input) and Hole 9 (b): the SHRED models have been trained using subsets of sensors from Hole 3. The black line is the experiments, the red one the CFD simulation and the dashed blue the SHRED prediction.}
    \label{fig: chap07-triga-exp-fromhole3}
\end{figure*}
At first, the case in which training CFD sensor measurements come from Hole 3 is considered. Once the SHRED models have been trained, experimental data from both available channels are used as input during the online phase: the evolution in time of the temperature field is plotted in Figure \ref{fig: chap07-triga-exp-fromhole3} for the experimental sensor locations. In Hole 3 (a), most of the experimental data are higher than the CFD counterpart, and SHRED tends to slightly increase the temperature in all sensor locations, implying that it detects that the background model is missing some information and thus it tries to recover it by increasing the temperature: this makes the prediction better from TC1 to TC6, but worse to TC7 and TC8: these two locations are the ones closer to the upper pool, which is not modelled by the CFD model and hence they are the locations where more difference between the background model and the experimental data are expected (in particular, due to recirculation phenomena that causes cold water from the pool to enter from the top of the core, the experimental data in these two points are lower than their CFD counterparts). For the temperature in Hole 9 (b), thus considering data unseen during training, the SHRED prediction is slightly increased compared to the CFD ones, for the reasons said above, providing an estimation of the temperature closer to the experimental measures. This behaviour can be justified by the fact that there is more discrepancy in Hole 9 between the experiments and the CFD simulation, and the SHRED recognizes this mismatch and tries to compensate by adjusting the overall magnitude. Overall, it seems that the SHRED architecture has some correction capabilities even for locations unseen during training.

\begin{figure*}[tp]
    \centering
    \begin{overpic}[width=0.95\linewidth]{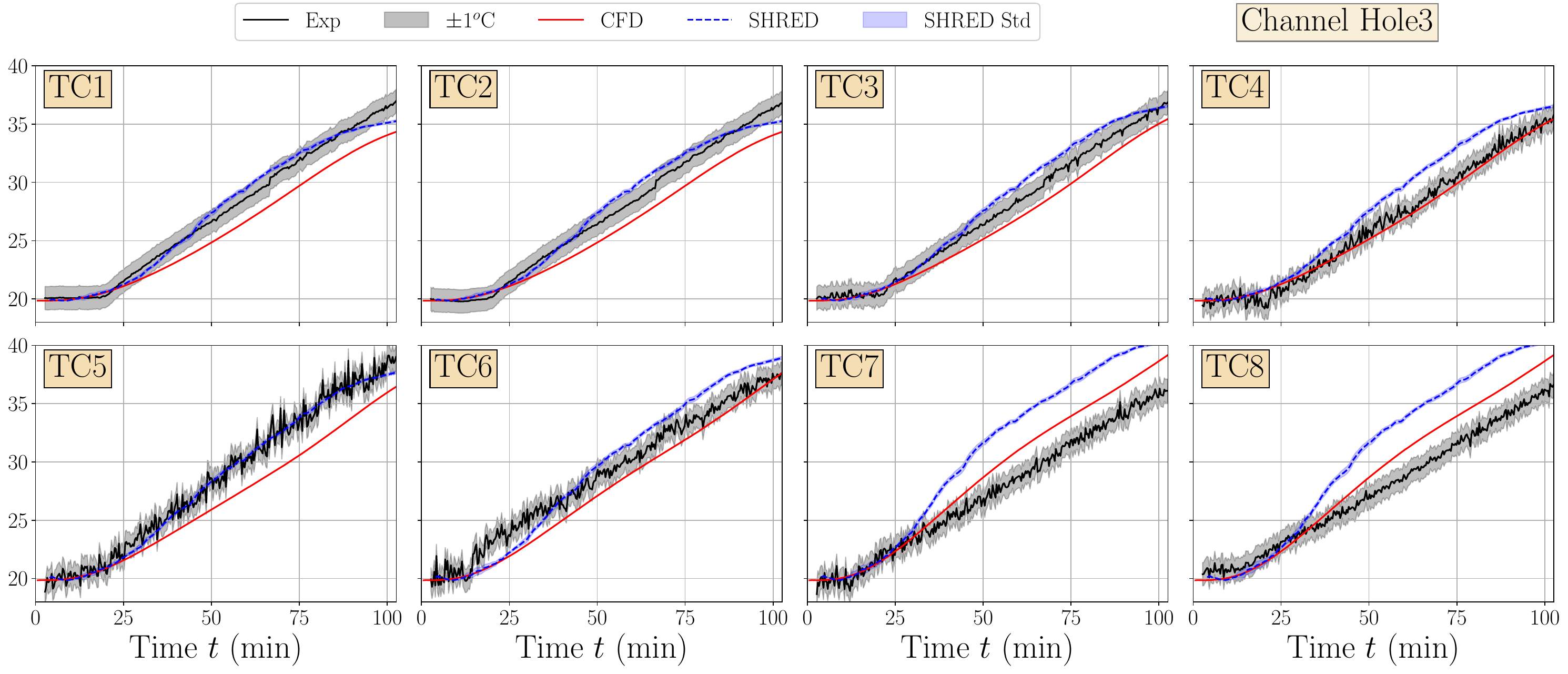}
    \put(95,41){(a)}   
    \end{overpic}\vspace{0.5cm}
    \begin{overpic}[width=0.95\linewidth]{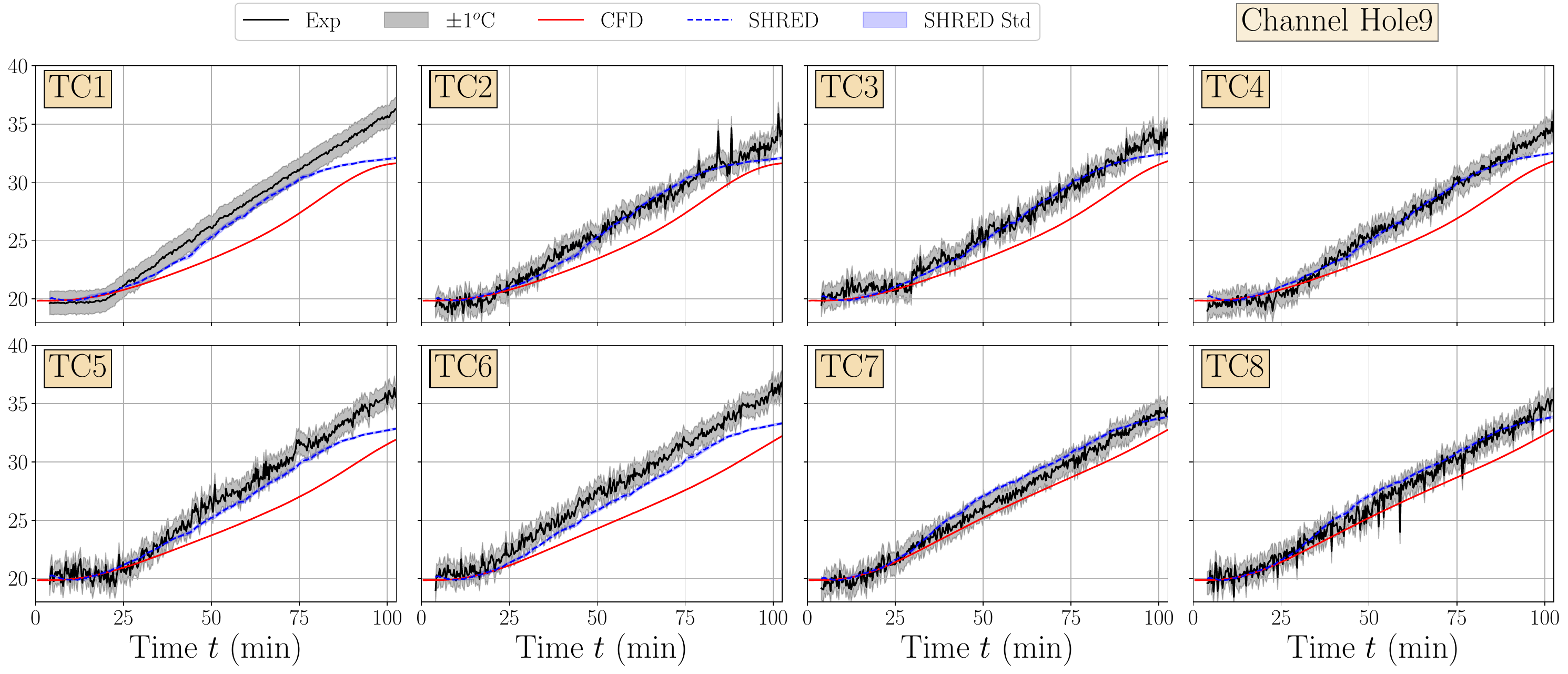}
    \put(95,41){(b)}   
    \end{overpic}
    \caption{Dynamical evolution of the temperature at the experimental locations in Hole 3 (a) and Hole 9 (b, input): the SHRED models have been trained using subsets of sensors from Hole 9. The black line is the experiments, the red one the CFD simulation and the dashed blue the SHRED prediction.}
    \label{fig: chap07-triga-exp-fromhole9}
\end{figure*}

Considering the case in which SHRED models are trained with sensors from Hole 9, reported in Figure \ref{fig: chap07-triga-exp-fromhole9}, a more significant update of the background model is observed for both Hole 3 (unobserved) and 9 (observed), compared to the previous case. In particular, there is a much better agreement between the SHRED prediction and the experimental data in almost every location, except TC7 and TC8 in Hole 3 (a) (likely for the same reasons as stated above, as these two locations are closer to the reactor pool), meaning that SHRED is not able to enforce the missing information (the presence of an upper pool and thus recirculation) of the background model in the updated prediction. The difference in correction between Hole 3 and 9 may be related to the greater accuracy of the CFD model in Hole 3 compared to  in Hole 9, and SHRED, in the former case, conceive the difference of the input trajectory as "noise" more than an actual discrepancy between model and data; on the other hand, for sensors in Hole 9 the CFD is less accurate, and SHRED updates the predictions accordingly, being able to recognize the presence of a discrepancy. In a real deployment of SHRED architectures for monitoring engineering system it is recommended to divide the available measurements into two sets: one used as input for SHRED, in ensemble mode for more robust predictions, and another to monitor the estimation during operation and eventually update the reduced model accordingly.

\begin{table*}[tp]
\centering
\small 
\begin{tabular}{lccccccc}
\toprule
 & \multicolumn{2}{c}{\textbf{Baseline CFD}} & \multicolumn{2}{c}{\textbf{SHRED (In. Hole 3)}} & \multicolumn{2}{c}{\textbf{SHRED (In. Hole 9)}} \\
\cmidrule(r){2-3} \cmidrule(r){4-5} \cmidrule(r){6-7}
 & \textbf{\textcolor{darkblue}{Hole 3}} & \textbf{\textcolor{darkorange}{Hole 9}} & \textbf{\textcolor{darkblue}{Hole 3}} & \textbf{\textcolor{darkorange}{Hole 9}} & \textbf{\textcolor{darkblue}{Hole 3}} & \textbf{\textcolor{darkorange}{Hole 9}} \\
\midrule
TC1  & \textcolor{darkblue}{1.697} & \textcolor{darkorange}{2.848} & \textcolor{darkblue}{1.318}          & \textcolor{darkorange}{2.559} & \textcolor{darkblue}{\textbf{0.612}} & \textcolor{darkorange}{\textbf{1.539}} \\
TC2  & \textcolor{darkblue}{1.541} & \textcolor{darkorange}{1.900} & \textcolor{darkblue}{1.172}          & \textcolor{darkorange}{1.604} & \textcolor{darkblue}{\textbf{0.710}} & \textcolor{darkorange}{\textbf{0.775}} \\
TC3  & \textcolor{darkblue}{1.215} & \textcolor{darkorange}{1.944} & \textcolor{darkblue}{\textbf{0.827}} & \textcolor{darkorange}{1.646} & \textcolor{darkblue}{0.917}          & \textcolor{darkorange}{\textbf{0.756}} \\
TC4  & \textcolor{darkblue}{0.590} & \textcolor{darkorange}{2.020} & \textcolor{darkblue}{\textbf{0.570}} & \textcolor{darkorange}{1.747} & \textcolor{darkblue}{1.736}          & \textcolor{darkorange}{\textbf{0.750}} \\
TC5  & \textcolor{darkblue}{2.290} & \textcolor{darkorange}{2.896} & \textcolor{darkblue}{1.877}          & \textcolor{darkorange}{2.596} & \textcolor{darkblue}{\textbf{0.737}} & \textcolor{darkorange}{\textbf{1.374}} \\
TC6  & \textcolor{darkblue}{1.495} & \textcolor{darkorange}{3.030} & \textcolor{darkblue}{\textbf{1.051}} & \textcolor{darkorange}{2.696} & \textcolor{darkblue}{1.535}          & \textcolor{darkorange}{\textbf{1.489}} \\
TC7  & \textcolor{darkblue}{\textbf{1.867}} & \textcolor{darkorange}{1.150} & \textcolor{darkblue}{2.322}          & \textcolor{darkorange}{0.898} & \textcolor{darkblue}{3.947}          & \textcolor{darkorange}{\textbf{0.743}} \\
TC8  & \textcolor{darkblue}{\textbf{1.806}} & \textcolor{darkorange}{1.449} & \textcolor{darkblue}{2.225}          & \textcolor{darkorange}{1.204} & \textcolor{darkblue}{3.868}          & \textcolor{darkorange}{\textbf{0.825}} \\
\midrule
Mean & \textcolor{darkblue}{1.563} & \textcolor{darkorange}{2.155} & \textcolor{darkblue}{\textbf{1.420}} & \textcolor{darkorange}{1.869} & \textcolor{darkblue}{1.758}          & \textcolor{darkorange}{\textbf{1.032}} \\
\bottomrule
\end{tabular}
\caption{Comparison of RMSE values with respect to the experimental thermocouples in Hole 3 (in blue) and Hole 9 (in orange). Bold styling distinguishes the optimal minimum error configuration per row within each channel.}
\label{tab: rmse_exp}
\end{table*}

\begin{figure*}[tp]
    \centering
    \includegraphics[width=0.83\linewidth]{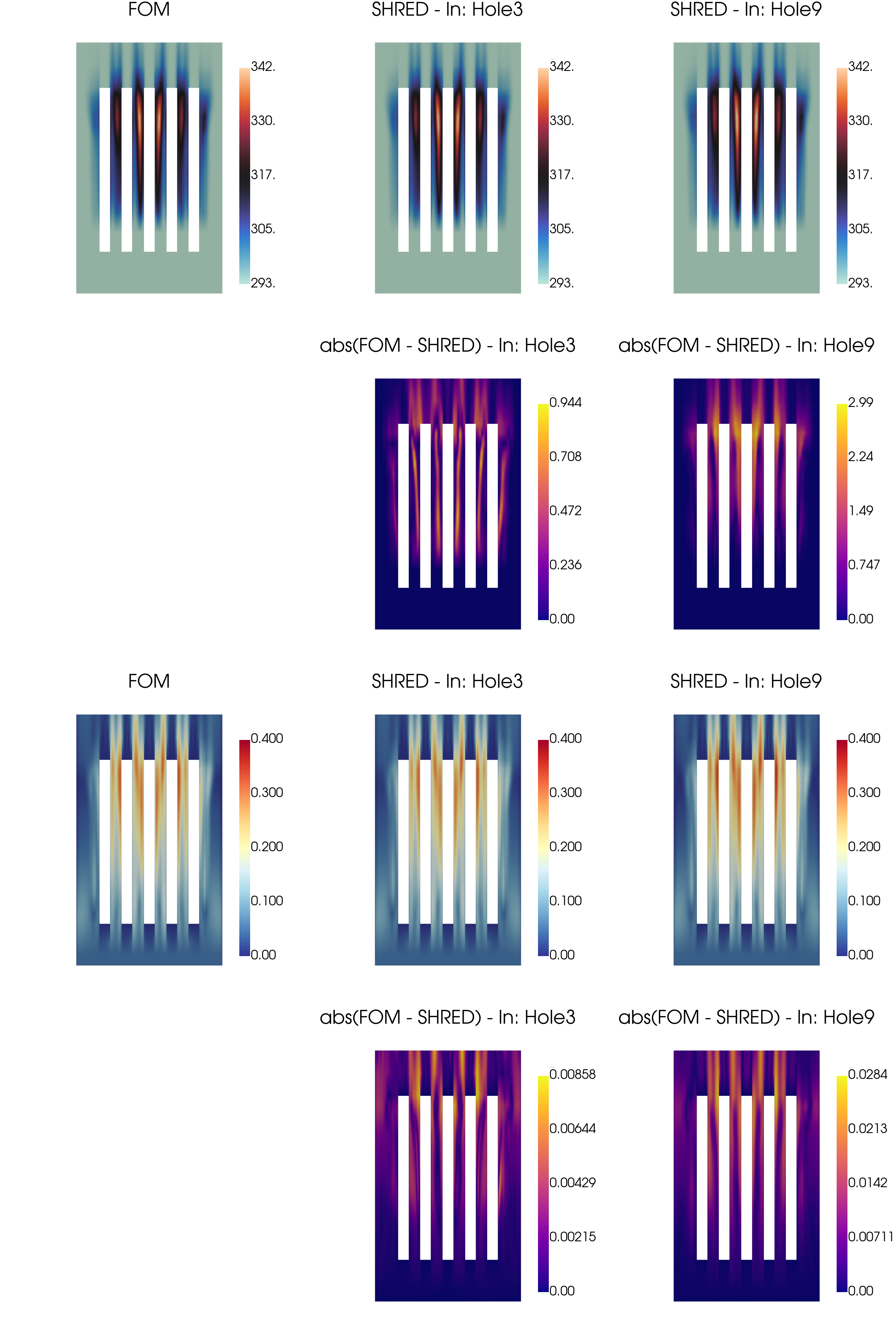}
    \caption{Contour plots for the last time instant in the test set of the temperature $T$ and velocity field $\vec{u}$: the section at $y=0$ of the reactor core is displayed. From left to right on the odd rows: full-order solution, the mean of the  SHRED ensemble for \textit{Hole 3} and \textit{Hole 9}. The even rows reports the residual fields.}
    \label{fig: shredrealdata-fom-contours}
\end{figure*}

For a more quantitative comparison, Table \ref{tab: rmse_exp} reports the Root Mean Square Error (RMSE) between the SHRED reconstruction or the CFD and the experimental data at the sensors locations in Hole 3 and 9. It clearly shows that in the channel where SHRED has sensor information the error decreases with respect to the CFD and on average the SHRED models achieve lower errors with respect to the simulation, apart from TC7 ad TC8 in Hole confirming the results of Figures \ref{fig: chap07-triga-exp-fromhole3} and \ref{fig: chap07-triga-exp-fromhole9}. In the end, Figure \ref{fig: shredrealdata-fom-contours} reports the reconstruction of the temperature and velocity field when real experimental data are adopted. It can be noted that the overall trend is similar to the CFD with only local magnitude correction as mentioned above. This fact shows that the network is indeed able to assimilate data "different" from the training ones, but it would require more advanced architectures to perform a complete data assimilation process \cite{bao2025dataassimilationdiscrepancymodeling, bao2026cheap2richmultifidelityframeworkdata}.

\section{Conclusions}\label{sec: conclusions}

This paper discusses the Shallow Recurrent Decoder architecture applied to the TRIGA Mark II research reactor, focusing in particular on the reconstruction capabilities of this neural architecture from sparse noisy measurements. This technique is able to map sparse sensors of a simple measurable quantity to the full state of the reactor, ensuring also outstanding capability in inferring unobservable fields. It has a great advantage compared to other methodologies of being agnostic to sensor positions, in the sense that its performance is generally not affected by their locations: from a practical standpoint, this means that the engineering constraints on sensors would not affect the reconstruction capability of the algorithm. 

This work investigates two different cases during the heating transient of the reactor from zero to full power in nominal operating conditions. At first, a verification of the approach is performed by considering only synthetic data randomly placed in two channels of the system, characterised by "low dynamics" (an external one and one shadowed by a control rod): despite this, SHRED is able to reproduce the system' state very accurately, and each reconstructed field has an average relative error below 3\%. Then, the use of real data is discussed, focusing on the possibility of updating the SHRED background knowledge using local observations after the deployment of the SHRED model. Real experimental measures collected on the reactor are available in two different channels: measures from one are used as input for training, whereas the ones from the other serve as validation of the approach; when the discrepancy between experimental measurements and CFD predictions is large enough to be clearly distinguished from noise, the SHRED architecture is able to rescale its output accordingly, improving the agreement with real data beyond what the background model alone provides. 

The present results are demonstrated for a single nominal operating scenario; the generalisation of the SHRED architecture to parametric or accidental conditions for the TRIGA reactor as done for the Molten Salt Reactor \cite{riva2025_parametricMSFR}, as well as a systematic comparison with alternative data-driven state estimation methods such as PBDW \cite{maday_parameterized-background_2014} and GEIM \cite{introini_stabilization_2023}, will be a topic for future investigation; including a more in-depth investigation of the SHRED as a operator learning architecture \cite{dang2025flronetdeepoperatorlearning}. More broadly, the lack of standardised benchmarks for scientific machine learning in nuclear engineering remains a recognised challenge, and structured efforts to address this are currently underway \cite{riva2026ctf4nuclearcommontaskframework}, with the authors actively contributing to this community effort.

This work shows that the SHRED architecture is able to correct, at least partially, the background knowledge of the model used for training; however, several limitations should be acknowledged. First, the present framework operates in reconstruction-only mode: no forecasting outside the observed time window is performed, which would be necessary for a full monitoring and control deployment. Second, the correction capability demonstrated in Section 4.2 is limited to a magnitude rescaling of the output, and does not constitute a complete data assimilation process in which the background model is updated in a physically consistent manner \cite{bao2025dataassimilationdiscrepancymodeling, bao2026cheap2richmultifidelityframeworkdata}. Third, the results are demonstrated for a single nominal operating scenario, and generalisation to parametric or accidental conditions remains an open question for the TRIGA reactor.

Future work will address these limitations along three main directions: (i) integration of SHRED within a true data assimilation framework \cite{bao2025dataassimilationdiscrepancymodeling, bao2026cheap2richmultifidelityframeworkdata}, in which model-to-data discrepancies are systematically corrected using incoming measurements; (ii) adoption of the SINDy-SHRED extension \cite{gao_sparse_2025}, 
which combines sparse measurement reconstruction with the discovery of interpretable\footnote{In the context of model discovery, interpretable is meant in a set of human-readable system of ordinary differential equations, relatively easy to solve.}, closed-form governing equations and enables forecasting of future states; and (iii) extension to parametric and off-nominal transients of the TRIGA system, as well as systematic benchmarking against alternative 
data-driven state estimation approaches \cite{maday_parameterized-background_2014, introini_stabilization_2023}.

\section*{Code and supplementary materials}  
The code and data (compressed) are available at: \href{github.com/ERMETE-Lab/NuSHRED}{github.com/ERMETE-Lab/NuSHRED}.

\section*{Acknowledgments} 
The contribution of Nathan Kutz was supported in part by the US National Science Foundation (NSF) AI Institute for Dynamical Systems (dynamicsai.org), grant 2112085.  JNK also acknowledges support from the Air Force Office of Scientific Research (FA9550-24-1-0141)

\section*{List of Symbols}
\nomenclature[L]{$p$}{Pressure}
\nomenclature[L]{$s$}{Number of input sensors for SHRED}
\nomenclature[L]{$r$}{Rank of the SVD/POD}
\nomenclature[L]{$t$}{Time}
\nomenclature[L]{$L$}{Lags}
\nomenclature[L]{$N_t$}{Number of time instants}
\nomenclature[L]{$T$}{Temperature}
\nomenclature[L]{$\vec{f}$}{Generic neural network function}
\nomenclature[L]{$\vec{u}$}{Discrete state vector/velocity field}
\nomenclature[L]{$\vec{v}$}{Reduced coefficients}
\nomenclature[L]{$\vec{x}$}{Space vector}
\nomenclature[L]{$\vec{y}$}{Measurement vector}
\nomenclature[L]{$\vec{y}^s$}{Input measurement vector for SHRED}
\nomenclature[L]{$\mathbb{H}$}{Observation operator}
\nomenclature[L]{$\mathbb{U}$}{Matrix with the SVD/POD modes}
\nomenclature[L]{$\mathbb{V}$}{Matrix with the SVD/POD reduced coefficients}
\nomenclature[L]{$\mathbb{X}$}{Snapshot matrix}
\nomenclature[L]{$\mathcal{J}$}{Loss function}
\nomenclature[L]{$\mathcal{L}$}{Number of ensemble models}
\nomenclature[L]{$\mathcal{N}_h$}{Dimension of the spatial grid}
\nomenclature[L]{$\mathcal{V}$}{Full state}

\nomenclature[G]{$\epsilon$}{Random noise}
\nomenclature[G]{$\kappa$}{Turbulent kinetic energy}
\nomenclature[G]{$\omega$}{Turbulent rate of dissipation}
\nomenclature[G]{$\psi$}{Generic field}
\nomenclature[G]{$\sigma$}{Standard deviation for random noise}
\nomenclature[G]{$\varepsilon_2^\psi$}{Average relative error in $l_2$ norm for the generic field $\psi$}
\nomenclature[G]{$\Omega$}{Spatial domain}
\nomenclature[G]{$\Xi^t$}{Subset of the time interval}

\nomenclature[A]{CFD}{Computational Fluid Dynamics}
\nomenclature[A]{DA}{Data Assimilation}
\nomenclature[A]{INFN}{Instituto Nazionale di Fisica Nucleare, National Institue of Nuclear Physics}
\nomenclature[A]{LENA}{Laboratorio di Energia Nucleare Applicata, Laboratory of Applied Nuclear Energy}
\nomenclature[A]{LSTM}{Long Short-Term Memory}
\nomenclature[A]{ML}{Machine Learning}
\nomenclature[A]{PDE}{Partial Differential Equation}
\nomenclature[A]{POD}{Proper Orthogonal Decomposition}
\nomenclature[A]{ROM}{Reduced Order Modelling}
\nomenclature[A]{SDN}{Shallow Decoder Network}
\nomenclature[A]{SHRED}{SHallow REcurrent Decoder}
\nomenclature[A]{SVD}{Singular Value Decomposition}
\nomenclature[A]{TC}{ThermoCouple}
\nomenclature[A]{TRIGA}{Training Research Isotope production General Atomic}
\footnotesize{ \printnomenclature }

\appendix
\section{The Full Order Model with OpenFOAM}\label{app: model}

Whereas the complete, transient multi-physics model of the TRIGA reactor core is currently under development, a CFD model in the OpenFOAM environment \cite{weller_tensorial_1998} was developed and validated by Introini et al. \cite{introini_advanced_2021, INTROINI2021111431, INTROINI2023112118}. To reduce the computational burden of the simulations, some simplifications and approximations were introduced: the overall fluid domain is reduced, and only the core is kept, the graphite reflector is cut, and its shape is approximated as a cylinder with the core in the middle, and among the various structures in the pool, only the guide tube for the control rods are considered due to their non-negligible influence on the flow (see Figure 7 from \cite{INTROINI2023112118}). The lower and upper grid represent challenging structures to be modelled (from a geometrical point of view); therefore, a porous medium approach \cite{Hafsteinsson2009_porousMediaOF} has been used \cite{introini_advanced_2021}.

In terms of modelling choices, the fluid domain contains incompressible water with constant physical properties. The buoyancy effects are taken into account adopting the Boussinesq approximation \cite{versteeg_introduction_2007}, in which the buoyancy forces depend linearly on the temperature field as $\rho \sim \vec{g}\beta(T-T_\infty)$, with $\vec{g}$ being the gravity acceleration, $\beta$ the thermal expansion coefficient and $T_\infty$ the reference temperature. Therefore, the governing equations are the incompressible Navier-Stokes equations coupled with the energy equation under the Boussinesq approximation:
\begin{equation}
    \left\{
        \begin{aligned}
            &\nabla\cdot \vec{u}=0\\
            &\dpart{\vec{u}}{t}+(\vec{u}\cdot\nabla)\vec{u}=\nu\Delta \vec{u}-\nabla p+\vec{g}\,\beta(T-T_\infty)\\
            &\dpart{T}{t}+\vec{u}\cdot \nabla T=\alpha \Delta T
        \end{aligned}
    \right.
\end{equation}
where $\vec{u}$ is the velocity vector, $\nu$ the kinematic viscosity, $p$ the normalised pressure to the density, $\vec{g}=[0, -g, 0]^T$ is the gravity acceleration, $T$ is the temperature field and $\alpha$ is the effective thermal diffusivity of the fluid. In terms of turbulence treatment, the model adopts a RANS approach with a $\kappa-\omega$SST model \cite{INTROINI2023112118}. 

The adopted simulation models only the thermal-hydraulics, thus the power produced by the fuel elements has to be introduced as a heat flux boundary condition: in particular, Cammi et al. \cite{CAMMI2016308} computed the discretised power distribution for each core ring using an MCNP model, which divides each fuel element into 8 sections to account for vertical ($\vec{e}_z$) axial variation. Through interpolation of the power as a function of height $z$, the model evaluates the sinusoidal heat flux distribution at the surface of each fuel element as $q'' = A\sin(Bz+C)+D$ in which $A,B,C,D$ are the regression coefficients that depend on the specific fuel element type and its position in the reactor core. A summary of all boundary conditions can be found in Table \ref{tab: bc-openfoam}.

The CFD model was developed in OpenFOAM 4 and solved using the \textit{buoyantBoussinesqPimpleFoam} solver for incompressible, turbulent flow under the Boussinesq approximation. The model counts 899954 mesh points, and the final mesh has been obtained following a grid independence study. The transient case refers to the heating of the reactor from zero-power condition and ambient temperature to steady-state operation at 250 kW, with the power given as a step insertion and considering an overall time interval $\mathcal{T}=[0,5]$ h for the generation of the synthetic data. Experimental measurements, instead, are available only for the first 100 minutes following the start of the transient: after this time, the temperature in the reactor pool exceeded the threshold value of 40 $^o$C, at which point the forced cooling, not included in the CFD model, starts. For more details on the CFD model and the experimental data, readers are referred to \cite{introini_advanced_2021, INTROINI2021111431, INTROINI2023112118}.

\begin{table}[htbp]
    \centering
    \begin{tabular}{lccc}
        \hline
        & Pool surface & Active zone & Other surfaces \\
        \hline
        Pressure & $p = 1 \ \text{atm}$ & $\frac{\partial p}{\partial \vec{n}} = 0$ & $\frac{\partial p}{\partial \vec{n}} = 0$ \\
        Velocity & $\frac{\partial \mathbf{u}}{\partial n} = 0$ & $\mathbf{u} = 0$ & $\mathbf{u} = 0$ \\
        Temperature & $T = T_R$ & $q'' = -k \frac{\partial T}{\partial \vec{n}}$ & $\frac{\partial T}{\partial \vec{n}} = 0$ \\
        Turbulence & $\frac{\partial}{\partial n} = 0$ & Wall functions & Wall functions \\
        \hline
    \end{tabular}
    \caption{Boundary conditions for the reactor model from \cite{INTROINI2023112118}.}
    \label{tab: bc-openfoam}
\end{table}

\section{Analysis of the Standard Deviation of the Ensemble Mode for SHRED}\label{sec: ensemble-std}

The authors proposed an ensembled strategy for SHRED models in \cite{riva2024robuststateestimationpartial}, able to provide more robust predictions especially with noisy measurements. Supposing to to have 10 available sensors: from these, different subsets (called configurations in the following) can be obtained and, since SHRED is cheap to train, it may be smart to train SHRED models with different configurations and then ensemble them to obtain a more robust state estimation \cite{riva2024robuststateestimationpartial}. Hence, the final prediction would be more statistically significant, as it will be characterised by a mean value and a standard deviation (hence a confidence interval for the true value). 

In order to predict the state at time $t_k$, the previous $k$ time instants are saved. For each sensor configuration $(l)$, the trajectories of the sensors are lagged as $\left[\vec{y}^{(l)}_{k}, \dots , \vec{y}^{(l)}_{k-K}\right]$ for $l=1, \dots, L$; the SHRED models are trained, producing $L$ outputs for the SVD coefficients $\vec{v}_k^{(l)}\in\mathbb{R}^r$ at time $t_k$. Recalling that the final goal consists in predicting the true SVD coefficients $\vec{v}_k\in\mathbb{R}^r$, at each time $t_k$, the SHRED output can be considered as samples from which an estimator for the mean can be obtained which is a random variable itself, with a Gaussian distribution following the central limit theorem
\begin{equation}
    \vec{v}_k \approx \hat{\vec{v}}_k \sim \mathcal{N}\left(\hat{\overline{\vec{v}}}_k; 
    \;
    \Sigma_{\vec{v}}^k\right)\qquad \text{ where } \hat{\overline{\vec{v}}}_k=\frac{1}{L}\sum_{l=1}^{L}\vec{v}
\end{equation}
where $\hat{\overline{\vec{v}}}_k$ is an estimator for the mean and $\Sigma_{\vec{v}}^k\in\mathbb{R}^{r\times r}$ is the covariance matrix, which is generally unknown and hence a proper unbiased estimator can be constructed
\begin{equation}
    \Sigma_{\vec{v}}^k\approx \hat{\Sigma}_{\vec{v}}^k = {\frac{1}{L-1}\sum_{l=1}^{L} \left(\vec{v}^{(l)}_k - \hat{\overline{\vec{v}}}_k\right)^2}
\end{equation}
which is non-diagonal in general. This accounts for the variances introduced by the sensor positions, the noise of the input measurements, and any uncertainty related to the training dataset. Adopting this approach, instead of having a single prediction, the ensemble strategy provides a more robust estimation of the state space, which is particularly useful in the presence of noisy data, allowing to improve the performance of SHRED and obtain more reliable results \cite{riva2024robuststateestimationpartial}.

\paragraph{Rescaling variance} From numerical experience, it is typically convenient to have the output predicted by SHRED in the range $[0,1]$ and hence the POD coefficients $\vec{v}_k$ are rescaled to a new set of vector with the following linear transformation
\begin{equation}
    \vec{b}_k = \mathbb{A}\left(\vec{v}_k - \vec{v}^{\text{min}}\right), \qquad
    \mathbb{A} = \text{diag}\left(\frac{1}{\vec{v}^{\text{max}} - \vec{v}^{\text{min}}}\right)
\end{equation}
where $\vec{b}_k$ are the rescaled POD coefficients, $\vec{v}^{\text{max}}$ and $\vec{v}^{\text{min}}$ are the maximum and minimum values of the POD coefficients over the training set. The inverse transformation provides:
\begin{equation}
\mathbf{b}(t; \boldsymbol{\mu}) = \mathbb{A}\cdot \mathbf{v}(t; \boldsymbol{\mu}) + \mathbf{b}_{min}
\end{equation}
Since the covariance does not scale linear, it must be rescaled properly following the rules of uncertainty propagation
\begin{equation}
\hat{\Sigma}_{\vec{v}}^k  = \mathbb{A}\cdot \hat{\Sigma}_{\vec{b}}^k  \cdot \mathbb{A}^T
\end{equation}

\paragraph{Decoding to the high-dimensional space.} The uncertainty can be then propagated to the high-dimensional state following a similar approach. Considering the more complete case in which the SHRED predicts the POD coefficients of multiple fields ($N_{\text{fields}}$)in a multi-physics context are elements of $\mathbb{R}^{r\cdot N_{\text{fields}}}$ (supposing the same rank $r$ for all of them for simplicity) and let $\vec{v}_k^{f_i}\in\mathbb{R}^r$ be the coefficients for field $f_i$ so that 
\begin{equation}
    \vec{v}_k = \left[\vec{v}_k^{f_1}, \vec{v}_k^{f_2}, \dots, \vec{v}_k^{f_{N_{\text{fields}}}} \right]^T
\end{equation}
The high-dimensional state of each field can be retrieved using the POD modes $\Phi_{i}\in\mathbb{R}^{\mathcal{N}_h\times r}$ using the linear combination $ \vec{u}_k^{f_i} = \Phi_{i}\vec{v}_k$. Let $\boldsymbol{\Phi}$ be a block diagonal matrix defined as follows
\begin{equation}
   \boldsymbol{\Phi} = \begin{bmatrix}
    \Phi_1 & 0 & \cdots & 0 \\
    0 & \Phi_2 & \cdots & 0 \\
    \vdots & \vdots & \ddots & \vdots \\
    0 & 0 & \cdots & \Phi_{N_{\text{fields}}}
    \end{bmatrix}
\end{equation}
such that the full high-dimensional state $\vec{U}\in\mathbb{R}^{\mathcal{N}_h\cdot N_{\text{fields}}}$ is retrieved as $\vec{U} = \boldsymbol{\Phi}\vec{v}_k$. The covariance matrix for the high-dimensional state (considering all the fields) can be estimated adopting the rules of uncertainty propagation, similar to the scaling process outlined above
\begin{equation}
\hat{\Sigma}_{\vec{U}}^k =  \boldsymbol{\Phi} \cdot \hat{\Sigma}_{\vec{v}}^k \cdot  \boldsymbol{\Phi}^T\in\mathbb{R}^{\mathcal{N}_h\cdot N_{\text{fields}}\times \mathcal{N}_h\cdot N_{\text{fields}}}
\end{equation}
which is generally too big to get stored in the memory. The block structure of $\boldsymbol{\Phi}$, each block of $\hat{\Sigma}_{\vec{U}}^k$ can be easily computed:
\begin{equation}
\hat{\Sigma}_{\vec{U}}^k =
\begin{bmatrix}
\Phi_1 \cdot \hat{\Sigma}_{\vec{v}}^k \cdot \Phi_1^T & \Phi_1 \cdot \hat{\Sigma}_{\vec{v}}^k \cdot \Phi_2^T & \cdots & \Phi_1 \cdot \hat{\Sigma}_{\vec{v}}^k \cdot \Phi_{N_{\text{fields}}}^T \\
\Phi_2 \cdot \hat{\Sigma}_{\vec{v}}^k\cdot \Phi_1^T & \Phi_2 \cdot \hat{\Sigma}_{\vec{v}}^k \cdot \Phi_2^T & \cdots & \Phi_2 \cdot \hat{\Sigma}_{\vec{v}}^k \cdot \Phi_{N_{\text{fields}}}^T \\
\vdots & \vdots & \ddots & \vdots \\
\Phi_{N_{\text{fields}}} \cdot \hat{\Sigma}_{\vec{v}}^k \cdot \Phi_1^T & \Phi_{N_{\text{fields}}} \cdot \hat{\Sigma}_{\vec{v}}^k\cdot \Phi_2^T & \cdots & \Phi_{N_{\text{fields}}} \cdot \hat{\Sigma}_{\vec{v}}^k \cdot \Phi_{N_{\text{fields}}}^T
\end{bmatrix}
\end{equation}
Since the standard deviation for field $f_i$ is typically the desired quantity, it is only sufficient to compute the diagonal of this huge matrix
\begin{equation}
\text{diag}\left(\hat{\Sigma}_{\vec{u}^{f_i}}^k\right) = \text{diag}\left(\Phi_i \cdot \hat{\Sigma}_{\vec{v}}^k \cdot \Phi_i^T\right) = \sum_{n=1}^{r} (\Phi_i)_{:,n}^2 \cdot (\hat{\Sigma}_{\vec{v}}^k)_{n,n}
\end{equation}
where $(\Phi_i)_{:,n}$ is the $n$-th column of the POD basis $\Phi_i$ for field $f_i$ and $(\hat{\Sigma}_{\vec{v}}^k)_{n,n}$ is the $n$-th diagonal element of the covariance matrix of the POD coefficients. Using \texttt{torch} operations, this can be efficiently computed as follows. Alternatively, this can be estimated directly from the high-dimensional reconstruction of each SHRED model, performing the ensemble at the full-order level rather than on the reduced one.

\bibliographystyle{unsrt}
\bibliography{bibliography}

\end{document}